\documentclass[a4paper, 11pt]{article}

\usepackage{jheppub}
\usepackage{abb}

\usepackage{amsmath, amsfonts}
\usepackage[normalem]{ulem}
\usepackage{bm}
\usepackage{physics}

\usepackage{comment}
\usepackage{color}
\usepackage{bm}

\makeatletter
\@addtoreset{equation}{section}

\makeatother

\allowdisplaybreaks[4]

\usepackage{cancel}

\usepackage{bbold}

\usepackage{empheq}
\usepackage{framed}
\usepackage[most]{tcolorbox}
\tcbuselibrary{breakable}
\usepackage{xcolor}
\colorlet{shadecolor}{orange!15}

\usepackage{caption}
\usepackage{subcaption}

\usepackage{tikz}
\usetikzlibrary{intersections, calc, arrows.meta}
\usetikzlibrary{patterns}
\usepackage{pgfplots}

\usepackage{pst-venn}

\usepackage{array}
\usepackage{longtable}
\usepackage{colortab}
\usepackage{colortbl}
\usepackage{arydshln}

\usepackage[symbol]{footmisc}

\newcommand{\ctext}[1]{\raise0.2ex\hbox{\textcircled{\scriptsize{#1}}}}

\title{
\boldmath Borel resummation of secular divergences in stochastic inflation
}

\author[a, b]{Masazumi Honda}
\author[c]{, Ryusuke Jinno}
\author[d]{, Lucas Pinol}
\author[c, e]{, and Koki Tokeshi}

\affiliation[a]{Center for Gravitational Physics and Quantum Information, Yukawa Institute for Theoretical Physics (YITP), Kyoto University, Sakyo, Kyoto 606-8502, Japan}
\affiliation[b]{RIKEN iTHEMS,
2-1 Hirosawa, Wako, Saitama 351-0198, Japan}
\affiliation[c]{Research Center for the Early Universe (RESCEU),
Graduate School of Science, The University of Tokyo, Tokyo, 113-0033, Japan}
\affiliation[d]{Instituto de F\'{i}sica T\'{e}orica UAM-CSIC, Calle Nicolás Cabrera 13-15, 28049,
Madrid, Spain}
\affiliation[e]{Graduate School of Science, The University of Tokyo, Bunkyo, Tokyo 113-0033, Japan}

\emailAdd{masazumi.honda@yukawa.kyoto-u.ac.jp}
\emailAdd{ryusuke.jinno@resceu.s.u-tokyo.ac.jp}
\emailAdd{lucas.pinol@ift.csic.es}
\emailAdd{tokeshi@resceu.s.u-tokyo.ac.jp}

\abstract{
We make use of Borel resummation to extract the exact time dependence from the divergent series found in the context of stochastic inflation.
Correlation functions of self-interacting scalar fields in de Sitter spacetime are known to develop secular IR divergences via loops, and the first terms of the divergent series have been consistently computed both with standard techniques for curved spacetime quantum field theory and within the framework of stochastic inflation.
We show that Borel resummation can be used to interpret the divergent series and to correctly infer the time evolution of the correlation functions.
In practice, we adopt a method called Borel--Pad\'{e} resummation where we approximate the Borel transformation by a Pad\'{e} approximant.
We also discuss the singularity structures of Borel transformations and mention possible applications to cosmology. 
}

\begin{document}
\begin{flushright}
\small
YITP-23-43, RIKEN-iTHEMS-Report-23, RESCEU-7/23
\end{flushright}
\maketitle
\flushbottom

\section{Introduction}
\label{sec:section1}

Cosmic inflation, an era of quasi de Sitter expansion of the early Universe, is now the leading paradigm to describe the earliest cosmological history. 
In addition to solving the horizon and flatness problems of the standard hot Big Bang model~\cite{10.1093/mnras/195.3.467, PhysRevD.23.347, Starobinsky:1980te, Linde:1981mu, PhysRevLett.48.1220, Linde:1983gd}, inflation provides an explanation for the origin of structures in our Universe. 
Vacuum fluctuations are generated deep inside the horizon and then stretched to cosmological scales by the accelerated expansion, thus seeding Cosmic Microwave Background (CMB) anisotropies and the Large-Scale Structure (LSS).
The simplest model of inflation, a slowly-rolling scalar field, predicts nearly scale-invariant, adiabatic, and sufficiently small cosmological scalar fluctuations, being consistent with the current large-scale observations~\cite{Planck2018VI, Planck2018X}.

Although the physics of inflation is rather constrained at the largest cosmological scales emerging from vacuum fluctuations during an epoch deep inside the inflationary era, much remains to be understood about the remaining of the inflationary evolution through small-scale observations.
For example, if fluctuations at small cosmological scales are sufficiently enhanced compared to those at the CMB ones, a significant amount of primordial black holes (PBHs) could be produced when primordial perturbations re-enter the Hubble horizon in the radiation- and matter-dominated eras~\cite{1967SvA....10..602Z, 10.1093/mnras/152.1.75, 10.1093/mnras/168.2.399, 1975ApJ...201....1C}.
In those scenarios, it may happen that fluctuations at small scales are so large that standard perturbation theory breaks down and that one needs a non-perturbative formalism to describe them\footnote{
    It has recently been advocated that a dramatic enhancement of small-scale fluctuations could even lead to the breakdown of cosmological perturbation theory at CMB scales, therefore putting into question the viability of single-field PBH formation scenarios~\cite{Kristiano:2022maq, Kristiano:2023scm}. This conclusion has been criticized in~\cite{Riotto:2023hoz, Riotto:2023gpm}, see also~\cite{Ando_2021, Inomata:2022yte, Choudhury:2023vuj, Choudhury:2023jlt, Choudhury:2023rks, Firouzjahi:2023aum, Motohashi:2023syh} for recent works tackling this issue.
    Although the point of this work is not to address the viability of these scenarios, we simply mention that the stochastic formalism precisely allows one to describe situations where perturbation theory breaks down, and that a breakdown of perturbation theory does not necessarily imply ruling out of the model.
}.
Stochastic inflation~\cite{Starobinsky:1986fx} precisely enables one to treat such large fluctuations in a non-perturbative way, and correctly infer the statistical properties of primordial fluctuations in this regime.
Actually, it is now believed that the formation of primordial black holes is mostly sensitive to very large over-densities, rather than an overall increase of the root mean square density.
These rare events, located in the tails of the distribution of fluctuations, cannot be described by the usual perturbation theory, even without an amplification mechanism at a specific scale.
For example, the stochastic formalism has been used to prove that \textit{exponential tails} can typically develop away from the center of the distribution, potentially leading to many more primordial black holes than a Gaussian distribution with the same power spectrum~\cite{Pattison_2017, Pattison_2021,Vennin:2020kng, Ezquiaga_2020, PhysRevLett.127.101302,Achucarro:2021pdh}.
Other than their implication in terms of primordial black holes, small-scale amplification mechanisms are also investigated for their potential to lead to secondary gravitational waves at horizon re-entry (see~\cite{Domenech:2021ztg} for a recent review), as well as to spectral distortions in the CMB at intermediates scales (see, \textit{e.g.}~\cite{Kogut:2019vqh} for a recent update on the status and prospects of these observables).

In the stochastic inflation framework (see the seminal works~\cite{STAROBINSKY1982175,Starobinsky:1986fx,NAMBU1988441,NAMBU1989240,Kandrup:1988sc,Nakao:1988yi,Nambu:1989uf,Mollerach:1990zf,Linde:1993xx,Starobinsky:1994bd}), the long-wavelength modes of the scalar field are driven by an effectively classical, yet stochastic dynamics.
The source of the stochasticity stems from the quantum nature of the vacuum fluctuations of this bosonic field.
When these microphysical fluctuations are stretched on super-Hubble scales, they join the infrared sector of the scalar field.
One can therefore see the coarse-grained, long-wavelength modes, as an open system subject to a constant interaction with a bath of ultra-violet modes.
This interaction is most notably described as a noise term in a stochastic differential equation for the infrared system, called a Langevin equation.
Correspondingly, from the Langevin equation (and given a time discretisation scheme), one can consider the associated Fokker--Planck equation for the probability density function (PDF) of coarse-grained modes. 
Although the dynamics may be exactly solvable for some classes of systems, it is in general difficult to obtain the full behaviour without numerical calculations.
However, important analytical results have been derived with the stochastic formalism, at least in three directions.
\begin{itemize}
    \item 
The first one, which is also the main focus of this work, is the series expansion of the field's correlation functions in terms of $\log a$, where $a$ is the scale factor.
Excellent agreement has been shown between the stochastic approach and various quantum field theory techniques, mostly in the paradigmatic setup of $\lambda \phi^4$ theory, for the first orders of the series expansion~\cite{Starobinsky:1994bd,Prokopec:2007ak,Finelli:2008zg,Finelli:2010sh,Garbrecht:2013coa,Garbrecht:2014dca,Onemli:2015pma,Cho:2015pwa} (see also~\cite{Pinol:2018euk,Pinol:2020cdp} for the first such predictions in the context of multifield stochastic inflation with curved field space).
As these time series are divergent, an effect known as secular divergences, the expansion is reputed trustworthy at early times only; in this work we precisely revisit this question and extend the validity of these results up to sufficiently late time. 
 \item The second direction concerns the opposite, late-time expansion of the correlation functions and the related PDF.
These results (see, \textit{e.g.}~\cite{Seery:2007we,
Enqvist:2008kt,
2009JCAP...05..021S,
Burgess:2009bs,
Seery:2010kh,
Gautier:2013aoa,
Guilleux:2015pma,
Gautier:2015pca,
Hardwick:2017fjo,
Markkanen:2017rvi,
LopezNacir:2019ord,
Gorbenko:2019rza,
Mirbabayi:2019qtx,
Adshead:2020ijf, 
Moreau:2020gib,
Cohen:2020php
}), are impressive as they somehow encompass the late-time resummation of the IR secular divergences previously mentioned.
\item A third direction, more directly related to cosmology and inflation, concerns the application of the stochastic formalism to the derivation of primordial correlation functions for the curvature perturbation.
The so-called stochastic $\delta N$-formalism, enables one to derive those statistics from the fluctuations of the duration of inflation in uncorrelated patches of the universe~\cite{Salopek:1990jq,Sasaki:1995aw,Sasaki:1998ug,Lyth:2004gb,Fujita:2013cna,Fujita:2014tja,Vennin:2015hra,Kawasaki:2015ppx,Assadullahi:2016gkk,Vennin:2016wnk}, proving notably useful for predicting the abundance of primordial black holes~\cite{Kawasaki:2015ppx,Pattison:2017mbe,Ezquiaga:2018gbw,Biagetti:2018pjj,Ezquiaga:2019ftu,Panagopoulos:2019ail,Achucarro:2021pdh}.
\end{itemize}
In this work, we bridge the gap between the early- and late-time expansions of the stochastic formalism (the first and second points), by providing a way to resum the IR secular divergences at any finite of infinite time, 

\textit{Borel resummation}~\cite{ASENS_1899_3_16__9_0} is one of the standard methods to resum formally divergent series.
It not only makes sense out of divergent series, 
but also gives us information on non-perturbative effects through analytic structures in the Borel plane
via \textit{resurgence} relations~\cite{SC_1977__17_1_A5_0}.\footnote{
See~\cite{Costin:1999798,Marino:2012zq,Dorigoni:2014hea,ANICETO20191,2014arXiv1405.0356S} for some reviews.
}
While Borel resummation and resurgence have long history of applications to quantum mechanics~\cite{Bender:1969si,Bender:1973rz,Balian:1978et,AIHPA_1983__39_3_211_0,ZinnJustin:2004ib,ZinnJustin:2004cg,Jentschura:2010zza,Jentschura:2011zza,Dunne:2013ada,Basar:2013eka,Dunne:2014bca,Escobar-Ruiz:2015nsa,Escobar-Ruiz:2015rfa,Misumi:2015dua,Behtash:2015zha,Behtash:2015loa,Gahramanov:2015yxk,Dunne:2016qix,Kozcaz:2016wvy,Fujimori:2016ljw,Dunne:2016jsr,Serone:2016qog,Basar:2017hpr,Alvarez:2017sza,Behtash:2018voa,Duan:2018dvj,Raman:2020sgw,Sueishi:2019xcj,Sueishi:2020rug,Sueishi:2021xti},
currently there are many applications to various other physics such as quantum field theory (QFT), hydrodynamics~\cite{Aniceto:2015mto,Basar:2015ava,Casalderrey-Solana:2017zyh,Behtash:2017wqg,Heller:2018qvh,Heller:2020uuy,Aniceto:2018uik,Behtash:2020vqk} and string theory~\cite{Marino:2008vx,Garoufalidis:2010ya,Chan:2010rw,Chan:2011dx,Schiappa:2013opa,Marino:2006hs,Marino:2007te,Marino:2008ya,Pasquetti:2009jg,Aniceto:2011nu,Santamaria:2013rua,Couso-Santamaria:2014iia,Grassi:2014cla,Couso-Santamaria:2015wga,Couso-Santamaria:2016vcc,Couso-Santamaria:2016vwq,Kuroki:2019ets,Kuroki:2020rgg,Dorigoni:2020oon}.
In particular, QFT recently has a variety of applications of Borel resummation and resurgence, including 2d QFTs~\cite{Dunne:2012ae,Dunne:2012zk,Cherman:2013yfa,Cherman:2014ofa,Misumi:2014jua,Nitta:2014vpa,Nitta:2015tua,Behtash:2015kna,Dunne:2015ywa,Buividovich:2015oju,Demulder:2016mja,Sulejmanpasic:2016llc,Okuyama:2018clk,Abbott:2020qnl,Abbott:2020mba,Ishikawa:2019tnw,Ishikawa:2020eht},
the Chern-Simons theory~\cite{Gukov:2016njj,Gang:2017hbs,Wu:2020dhl,Fuji:2020ltq,Ferrari:2020avq,Gukov:2019mnk,Garoufalidis:2020nut},
4d non-supersymmetric QFTs~\cite{Argyres:2012vv,Dunne:2015eoa,Yamazaki:2017ulc,Mera:2018qte,Itou:2018wkm,Canfora:2018clt,Ashie:2019cmy,Ishikawa:2019oga,Unsal:2020yeh,Ashie:2020bvw,Morikawa:2020agf},
and supersymmetric gauge theories in diverse dimensions~\cite{Russo:2012kj,Aniceto:2014hoa,Aniceto:2015rua,Honda:2016mvg,Honda:2016vmv,Gukov:2016tnp,Honda:2017qdb,Gukov:2017kmk,Dorigoni:2017smz,Honda:2017cnz,Fujimori:2018nvz,Grassi:2019coc,Dorigoni:2019kux,Dorigoni:2021guq,Fujimori:2021oqg}.
However, there are only few applications so far in the cosmological or astrophysical contexts, and mainly for quasi-normal modes of a black hole~\cite{Hatsuda:2021gtn,Hatsuda:2019eoj,Matyjasek:2019eeu,Eniceicu:2019npi}.
The aim of the present paper is to present a new cosmological application of Borel resummation; 
from a truncated series at some finite order, 
we reconstruct the long-time evolution of the correlation function from transient to equilibrium regimes.
To make the setup as simple as possible, we mainly restrict ourselves to a spectator field in a quartic potential.
In order to make contrast with exactly solvable systems, we also discuss a spectator in a quadratic potential in a parallel way.

The organization of the paper is as follows. 
In Sec.~\ref{sec:section2}, we review the framework of stochastic inflation, focusing on the distribution and correlation functions of a test (spectator) scalar field in the presence of a quadratic or a quartic potential. 
There we perform a perturbative expansion of the correlation functions and see how it leads to a divergent behaviour in the $\lambda\phi^4$ theory.
In Sec.~\ref{sec:section3} we introduce Pad\'{e} approximants and Borel resummation 
to recover the correct behaviour of the correlation functions in time.
In the application of the Borel resummation, we specifically use a method called \textit{Borel--Pad\'{e} resummation} where we approximate Borel transformation (rather than the correlation functions themselves) by the Pad\'{e} approximant.
Section~\ref{sec:section4} is devoted to discussion and conclusions.
In Appendix~\ref{app:appendix1}, we discuss technicalities of the Borel--Pad\'{e} resummation technique. 
 
\section{Stochastic spectator, its PDF, and the Fokker--Planck equation}
\label{sec:section2}

The stochastic formalism for inflation~\cite{Starobinsky:1986fx} aims at dealing directly with the super-Hubble part of the quantum fields present during inflation.
It is derived as an effective field theory for the long-wavelength modes of scalar fields, after the short-wavelength modes have been integrated out.
The quantum properties of these small-scales degrees of freedom are imprinted in the statistical properties of a stochastic noise.
This noise then acts as a driving force on the effectively classical dynamics of the so-called coarse-grained fields on super-Hubble scales (see, \textit{e.g.} Refs.~\cite{Polarski:1995jg, Lesgourgues:1996jc, Polarski:2001yn, Kiefer:2008ku, Burgess:2014eoa, Martin:2015qta}, about the quantum-to-classical transition during inflation).
The resulting Langevin equations can then be translated into the Fokker--Planck equation, which describes the convection-diffusion of the probability distribution function for the coarse-grained fields.
The convection term is given by the usual background dynamics of the fields, and is often dictated by the derivative of a scalar potential (see also Refs.~\cite{Pinol:2018euk, Pinol:2020cdp} for the incorporation of non-minimal kinetic couplings between scalar fields in the context of stochastic inflation).
The diffusion term comes from the noise in the Langevin equation and describes the effect of the small-scale quantum modes crossing the cut-off scale and joining the open system made of super-Hubble fields.
In the following, in order to keep the discussion simple, we adopt the simpler approach of stochastic inflation from the point of view of the equations of motion.
The stochastic formalism can also be found from the theoretically robust path integral derivation, see Refs.~\cite{Morikawa:1989xz,Calzetta:1999zr,Matarrese:2003ye,Levasseur:2013ffa,Moss:2016uix,Tokuda:2017fdh,Prokopec:2017vxx,Pinol:2020cdp}.

Throughout this paper, we consider the dynamics of a spectator field $\phi$ during inflation.
In practice, we will work at leading order in the slow-roll parameters, which amounts to approximating the quasi-de Sitter background as an exact de Sitter one with a constant Hubble parameter $H = \dot{a} / a$, maintained by another scalar field playing the role of the inflaton.\footnote{
    One may think naively that the next-to-leading order behaviour taking into account corrections from a time-dependent Hubble scale could be obtained in an adiabatic way by replacing $H \rightarrow H(N)$ in equilibrium distributions and correlation functions.
    However, this was shown to be generally wrong in~\cite{Hardwick:2017fjo}, where it was explicitly proved that the time scale for spectator fields to relax to the equilibrium is typically much larger than the time scale of evolution of $H(N)$ (see also Ref.~\cite{Enqvist:2012xn}).
    Therefore, spectator fields are typically out of equilibrium during inflation, which actually provides a further motivation for the current work.
    We plan to address the situation of a more realistic inflationary background with non-adiabatic evolution of the spectator field in a future publication.
} 
Here, a dot means a derivative with respect to the cosmic time $t$, and $a = a (t)$ is the scale factor with exponential time-dependence.
In the following, rather than the cosmic time, we will use the convenient and deterministic (see Ref.~\cite{Finelli:2008zg}) variable $N=\log a$ called the number of $e$-folds, as a time variable.
We decompose the scalar field $\phi$ into UV modes ($ \phi_{>}$ for $k>k_\sigma(N)$) and IR ones ($ \phi_{<}$ for $k<k_\sigma(N)$), as 
\begin{equation}
  \phi (N, \vb*{x}) 
  = \underbrace{
    \int \frac{\dd^3 k}{(2 \pi)^3} \, \Theta ( k - k_{\sigma}(N) ) 
    \tilde{\phi} (N, \vb*{k}) e^{i \vb*{k} \cdot \vb*{x}} }_{\equiv \phi_{>} (N, \vb*{x})} 
  + \underbrace{
    \int \frac{\dd^3 k}{(2 \pi)^3} \, \Theta ( - k + k_{\sigma}(N) ) 
    \tilde{\phi} (N, \vb*{k}) e^{i \vb*{k} \cdot \vb*{x}} 
  }_{\equiv \phi_{<} (N, \vb*{x})} 
  \,\, . 
  \label{eq:stochinf_decomp}
\end{equation}
We introduced a time-dependent cut-off $k_{\sigma}(N)\equiv \sigma a(N) H$,
with $\sigma \ll 1$ a bookkeeping parameter representing the ratio between the physical size of the Hubble radius and the cut-off length.
Physically, it is chosen such that modes with wavelength larger than the cut-off scale can be well approximated as classical random variables, rather than fully quantum operators.
The fact that the cut-off is time dependent is crucial as time derivatives of the full field $\phi$ will also hit the window function $\Theta$, giving rise to terms absent in the corresponding deterministic, background theory (for which $k_\sigma(N) \rightarrow 0$).
Here we also defined $\Theta$ as the Heaviside step function, which amounts to a hard cut-off separating the UV sector from the IR one.
The choice of the window function is not irrelevant, as our choice of a hard cut-off will result in a white noise, while a smooth window function would have resulted in a colored noise with different statistical properties, see~\cite{Winitzki:1999ve,Matarrese:2003ye,Liguori:2004fa}.

The dynamics of the full field $\phi$, before the decomposition into IR and UV modes, is described by the Klein-Gordon equation 
\begin{equation}
  \pdv[2]{\phi}{N} + 3  \pdv{\phi}{N} - \frac{\grad^2 \phi}{a^2H^2} + \frac{1}{H^2}\dv{V}{\phi} = 0 \,\, , 
  \label{eq:stochinf_kg}
\end{equation}
where $V = V (\phi)$ is the scalar potential.
Inserting the decomposition~\eqref{eq:stochinf_decomp} into Eq.~\eqref{eq:stochinf_kg}, and assuming that the quantum fluctuations $\phi_{>}$ behave as in the usual linear perturbation theory, one finds the Langevin equation for the coarse-grained fields:
\begin{equation}
  \pdv{\phi}{N} 
  = - \frac{1}{3 H^2} \dv{V}{\phi} + \xi \,\, . 
  \label{eq:stochinf_lg}
\end{equation}
Here and in the following we simply write the long-wavelength field $\phi_{<}$ as $\phi$ since the stochastic formalism gives an effective description of $\phi_{<}$ only.
We have also assumed an overdamped approximation for the dyamics of the IR fields, that is that the acceleration term is negligible compared to the other ones.
This approximation is well motivated in situations where the scalar field is (at the classical, deterministic level), slowly rolling down the slope of its potential (see~\cite{Nakao:1988yi,Habib:1992ci} for the first works on stochastic inflation beyond slow roll).
The first term in the right hand side of Eq.~(\ref{eq:stochinf_lg}) describes the effect of the classical drift, while the second is the classical noise of quantum micro-physical origin, with correlation properties
\begin{equation}
  \expval{ \xi (N, \vb*{x}) \xi (N', \vb*{x}') } 
  = \frac{H^2}{4 \pi^2} \, \mathrm{sinc} (k_{\sigma} r) \, \delta_{\mathrm{D}} (N-N') \,\, .
  \label{eq:stochinf_noise}
\end{equation}
Here $r \equiv \abs{ \vb*{x} - \vb*{x}' }$ is the comoving distance between the two points.
In the following, we will be interested only in the one-point statistics of the fields, and therefore focus effectively on $r=0$ (and therefore $\mathrm{sinc}(k_\sigma r)\rightarrow 1$), although in practice our results will be more generally valid for any two points within a patch of the early universe with comoving size $r < k_\sigma^{-1}$.
We can interpret the presence of the Dirac $\delta$-distribution in time, $\delta_{\mathrm{D}}(N-N')$, as the fact that the stochastic dynamics is derived from a white noise.
The amplitude of the noise corresponds in general to the power spectrum of the quantum fluctuations $\phi_{>}$ when they cross the cut-off scale and correspond to the transfer of energy from the UV sector to the IR one.
In Eq.~\eqref{eq:stochinf_noise}, we have assumed that those fluctuations were behaving as being effectively massless, which yields a spectrum of amplitude $[H/(2\pi)]^2$.
In practice our amplitude for the noise can be thought of as the leading-order term in an expansion in $m_\mathrm{eff}^2/H^2$, where $m_\mathrm{eff}$ is the effective mass of the fluctuations $\phi_{>}$ at horizon crossing.
A technical assumption of this formalism is therefore that no non-perturbative mass will develop due to stochastic effects (see~\cite{Tokuda:2017fdh} for the first-order correction taking into account the backreaction of the stochastic dynamics on the amplitude of the noise, through the development of a mass term due to stochasticity).
It is also important to note the independence on $\sigma$ of the final Langevin equations for massless fields, at leading order and in the limit of $\sigma \ll 1$, see, \textit{e.g.} Refs.~\cite{Grain:2017dqa, Pinol:2020cdp, Ballesteros:2020sre} for discussions about a realistic choice of $\sigma$ for light --- but not massless --- fields.

The Langevin equation (\ref{eq:stochinf_lg}) can be translated into 
the Fokker--Planck equation for the probability density function (PDF) 
of $\phi$, $f(\phi, N)$, as
\footnote{In general, going from stochastic differential equations as the Langevin equations~\eqref{eq:stochinf_lg} to the Fokker--Planck equation, is far from trivial.
Indeed, when the amplitude of the noise is a function of the stochastic variable --- here $\phi$ ---, a situation called multiplicative noise, the stochastic dynamics depends on the discretisation time scheme in the Langevin equations, or equivalently, in the path integral representation of the theory.
The choice of time discretisation has been argued to exceed the accuracy of the stochastic formalism~\cite{Vennin:2015hra}, a statement proved to be correct for single-field inflation but resulting in an ambiguity dubbed ``inflationary stochastic anomalies'' and particularly relevant in multifield scenarios in~\cite{Pinol:2018euk}.
Based on a fundamental description at the level of the discretised path integral approach, it was proven that only the so-called Stratonovich scheme, corresponding to a mid-point discretisation, was leading to field-covariant equations in the stochastic formalism~\cite{Pinol:2020cdp}.
For the sake of this paper, the discretisation scheme is irrelevant as our stochastic variable, $\phi$, is a spectator field and the amplitude of the noise, given by $[H/(2\pi)]^2$, is independent of it.
}
\begin{equation}
  \pdv{f}{N} = \frac{1}{3 H^2} \pdv{\phi} \qty( \dv{V}{\phi} f ) + \frac{H^2}{8 \pi^2} \pdv[2]{f}{\phi} \,\, , 
  \qquad 
  f = f (\phi, \, N) \,\, .
  \label{eq:stochinf_fpeq}
\end{equation}
The \textit{stationary} solution of Eq.~(\ref{eq:stochinf_fpeq}), $\partial f_{\infty}  / \partial N = 0$, can be obtained for an arbitrary potential~\cite{PhysRevD.50.6357}, 
\begin{equation}
  f_{\infty} (\phi) 
  \equiv \lim_{N \to \infty} f (\phi, N) 
  = C \exp \qty[ - \frac{8 \pi^2}{3 H^4} V (\phi) ] \,\, , 
  \qquad 
  C^{-1} \equiv \int \dd \phi \, \exp \qty[ - \frac{ 8 \pi^2 }{3 H^4} V (\phi) ] \,\, . 
  \label{eq:stochinf_stat}
\end{equation}
For the initial condition, we assume that $\phi$ is deterministically located at a local minimum of $V (\phi$), 
$\phi = \phi_0$, at $N = 0$, 
\begin{equation}
  f_0 (\phi) 
  \equiv f (\phi, \, N = 0) 
  = \delta_{\mathrm{D}} (\phi - \phi_0) \,\, . 
  \label{eq:stochinf_inicond}
\end{equation}
Starting from $\phi_0$, the spectator field evolves according to Eq.~(\ref{eq:stochinf_lg}) 
with the classical drift and the quantum noise. 
As time passes by, the distribution of $\phi$ equilibrates to 
the stationary distribution given by Eq.~(\ref{eq:stochinf_stat}). 
The stationary correlation functions read
\begin{equation}
  \expval{ \phi^n }_{\infty}
  \equiv \lim_{N \to \infty} \expval{ \phi^n } (N) 
  = \int \dd \phi \, \phi^n f_{\infty} (\phi) \,\, . 
  \label{eq:stochinf_statexpv}
\end{equation}
While the equilibrium distribution and correlation functions are in general easy to obtain, it is often challenging to calculate the time evolution of the PDF and the correlations of $\phi$ without the help of numerical calculations.
In order to study the time evolution of the correlators in an analytic way, we expand them in terms of $N$, 
\begin{equation}
  \expval{ \phi^n } (N) 
  \equiv \int \dd \phi \, \phi^n f (\phi, \, N) 
  = \sum_{k = 0}^{\infty} a_{n, k} N^k \,\, . 
  \label{eq:stochinf_expv}
\end{equation}
Note that the right hand side is a formal perturbative series and is not guaranteed to converge.
Also, the coefficients $a_{n, k}$ have a mass dimension $n$ for any value of $k$.
The correlation functions $\expval{ \phi^n }$ can be shown to verify recurrence relations by using the integral definition Eq.~\eqref{eq:stochinf_expv} together with the Fokker--Planck equation~(\ref{eq:stochinf_fpeq}), after integration by parts:
\begin{align}
  \pdv{ \expval{ \phi} }{N} 
  &= - \frac{1}{3 H^2}\expval{ \dv{V}{\phi} } \,\, , \nonumber \\
  \pdv{ \expval{ \phi^n } }{N} 
  &= - \frac{1}{3 H^2} n \expval{ \dv{V}{\phi} \phi^{n-1} } + \frac{H^2}{8 \pi^2} n (n-1) \expval{ \phi^{n-2} } \quad \text{for} \quad n \geq 2 \,\, . 
  \label{eq:stochinf_expvrec}
\end{align}
One can already anticipate the difficulty about recovering exact expressions for $\expval{ \phi^n }$: Eq.~(\ref{eq:stochinf_expvrec}) may not represent a closed system of differential equations, depending on the choice of the scalar potential $V(\phi)$.
For the coefficients $a_{n, k}$, the initial and boundary conditions are set as follows.
From $\expval{ \phi^0 } = 1$, 
we should set $a_{0, k} = \delta_{0 k}$ for $k \in \mathbb{Z}_{\geq 0}$ with $\delta_{ij}$ being the Kronecker delta. 
We also set $a_{n, 0} = \delta_{n 0}$ for $n \in \mathbb{Z}_{\geq 0}$ from the deterministic initial condition $\phi=\phi_0$ (we consider $\phi_0 = 0$ for simplicity).
We will also assume here that the system has a $\mathbb{Z}_2$-symmetry, $\phi \leftrightarrow - \phi$, which further sets $a_{n = 2 m + 1, k} = 0$ for $m \in \mathbb{Z}_{\geq 0}$.
These assumptions are only technical and our formalism can also be applied to more diverse potentials and initial conditions.
Once the potential $V (\phi)$ is specified, and as long as it is a polynomial of a finite order, the coefficients $a_{n, k}$ can be recursively obtained from these conditions and Eq.~(\ref{eq:stochinf_expvrec}), as we will see through two specific examples below. 
\begin{figure}
  \begin{minipage}[b]{0.495\linewidth}
    \centering
    \subcaption{
        PDF in time for $V (\phi) = m^2 \phi^2 / 2$.
    }
    \includegraphics[width = 0.95\linewidth]{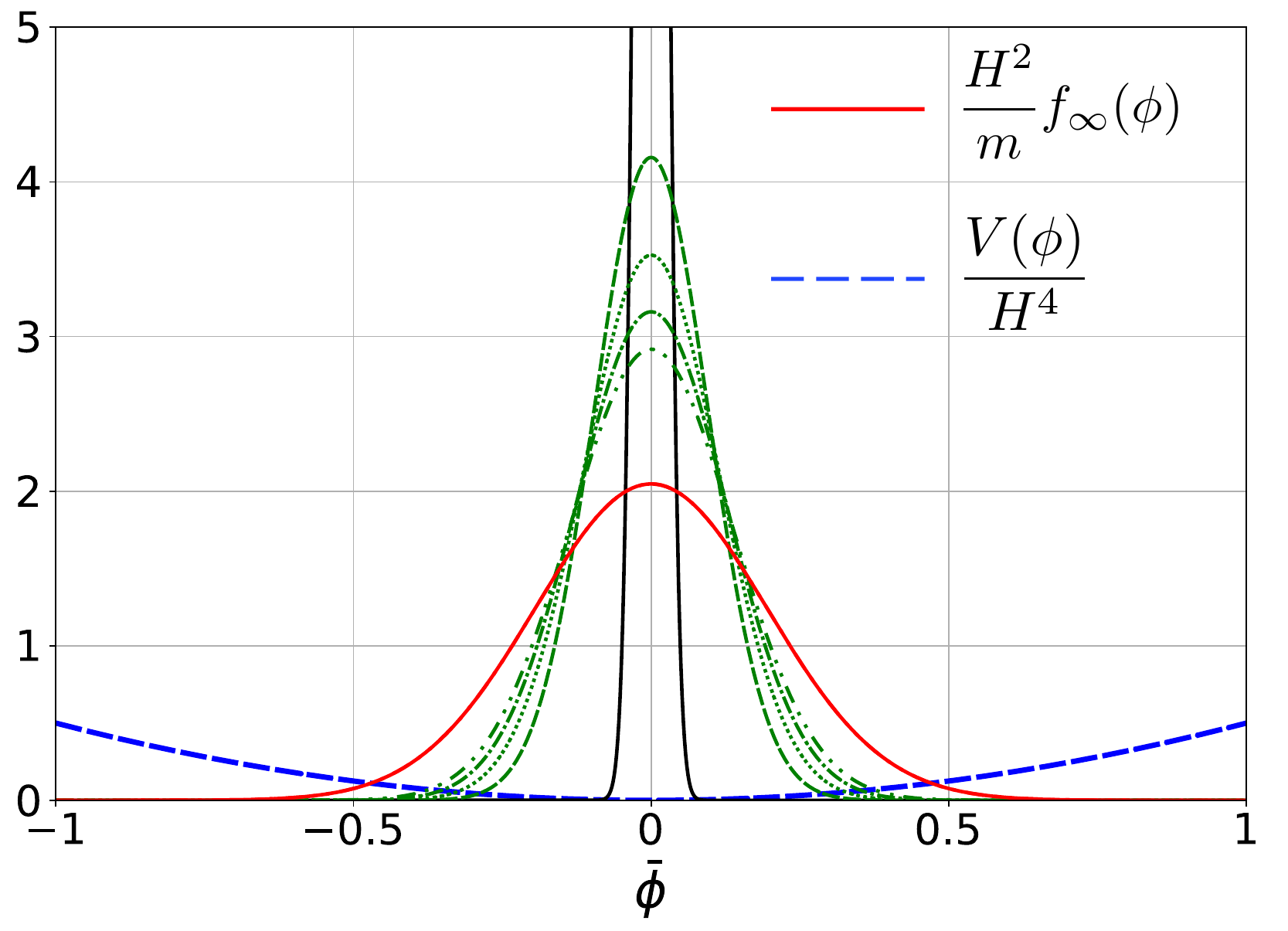}
  \end{minipage}
  \begin{minipage}[b]{0.495\linewidth}
    \centering
    \subcaption{
        PDF in time for $V (\phi) = \lambda \phi^4 / 4$.
    }
    \includegraphics[width = 0.95\linewidth]{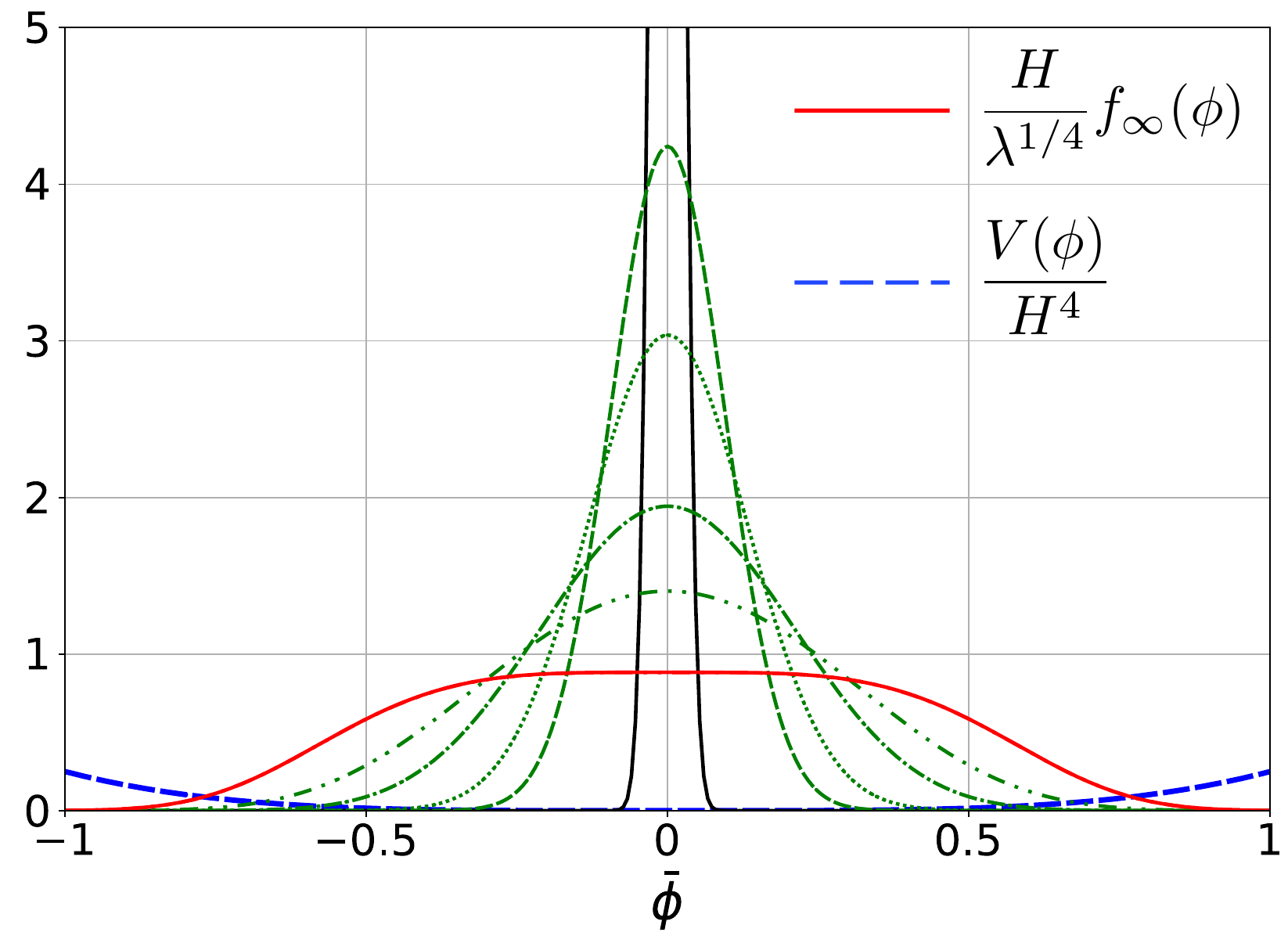}
  \end{minipage}
  \caption{
    Time evolution of the PDFs for $V (\phi) = m^2 \phi^2 / 2$ 
        (\textit{left}) and $V (\phi) = \lambda \phi^4 / 4$ (\textit{right}). 
        The black, red, and green lines represent respectively the initial, stationary, and transient distributions.
        The blue lines correspond to the scalar potentials. 
        The narrow Gaussian function, $f_0 (\phi) = e^{- \phi^2 / 2 \sigma^2} / \sqrt{2 \pi} \, \sigma$, is used as the initial distribution where $\sigma = 0.02$, in order to mimick the idealized $\delta$-function, Eq.~(\ref{eq:stochinf_inicond}). 
        Here, the Crank--Nicolson scheme was used to numerically solve the Fokker--Planck equation~(\ref{eq:stochinf_fpeq}).  
  }
  \label{fig:eqb_pdf}
\end{figure}
Throughout this paper, we consider a scalar field in a quadratic or quartic potential,
\begin{equation}
  V (\phi) 
  = \begin{cases}
    \displaystyle \frac{m^2}{2} \phi^2 & (\text{quadratic}) \,\, , \\[2.0ex] 
    \displaystyle \,\, \frac{\lambda}{4} \,\, \phi^4 & (\text{quartic}) \,\, . 
  \end{cases}
  \label{eq:stochinf_pot}
\end{equation}
Figure \ref{fig:eqb_pdf} shows the time evolution of the PDFs obtained by a numerical resolution of the Fokker--Planck equation~\eqref{eq:stochinf_fpeq} for each potential.
The stationary PDFs are obtained from Eq.~(\ref{eq:stochinf_stat}), 
\begin{equation}
  f_{\infty} (\phi) 
  = \begin{cases}
    \displaystyle \frac{m}{H^2} \sqrt{ \frac{4 \pi }{3} } \exp \qty[ - \frac{4 \pi^2 }{3} \qty( \frac{m}{H^2} \phi )^2 ] & (\text{quadratic}) \,\, , \\[2.0ex] 
    \displaystyle \frac{ \lambda^{1/4} }{H} \frac{2}{ \Gamma (1/4) } \qty( \frac{2 \pi^2}{3} )^{1/4} \exp \qty[ - \frac{2 \pi^2}{3 } \qty( \frac{ \lambda^{1/4} }{H} \phi )^4 ] & (\text{quartic}) \,\, . 
  \end{cases}
\end{equation}
For $V (\phi) = m^2 \phi^2 / 2$, the system conserves at late times an exactly Gaussian behaviour (actually, the distribution can be shown analytically to be Gaussian at any time, and the time-dependent standard deviation can be computed exactly), while for $V (\phi) = \lambda \phi^4 / 4$ a negative kurtosis develops. 
The two-point function at equilibrium reads
\begin{equation}
  \expval{ \phi^2 }_{\infty}
  \equiv \lim_{N \to \infty} \expval{ \phi^2 } (N)  
  = \begin{cases}
    \displaystyle \frac{ 3 }{8 \pi^2 } \qty( \frac{ H^2 }{m} )^2  & (\text{quadratic}) \,\,, \\[2.0ex] 
    \displaystyle \sqrt{ \frac{3}{2 \pi^2} } \frac{ \Gamma (3/4) }{ \Gamma (1/4) } \qty( \frac{H}{\lambda^{1/4}} )^{2} & (\text{quartic}) \,\, , 
  \end{cases}
  \label{eq:stochinf_expveq}
\end{equation}
and 
\begin{equation}
    \expval{ \phi^4 }_{\infty}
  = \begin{cases}
    \displaystyle \frac{27}{64 \pi^4} \qty( \frac{ H^2 }{m} )^4  & (\text{quadratic}) \,\,, \\[2.0ex] 
    \displaystyle \frac{3}{8 \pi^2} \qty( \frac{H}{\lambda^{1/4}} )^4& (\text{quartic}) \,\, . 
  \end{cases}
  \label{eq:stochinf_expveq_4pt}
\end{equation}
It is also possible to compute analytically the higher-order stationary correlation functions, \textit{e.g.} in the quartic case one can compute the kurtosis, $\expval{ \phi^4 }_{\infty} / \expval{\phi^2}_{\infty}^2 - 3 \approx - 0.812$, showing that the $\lambda \phi^4$ theory is \textit{platykurtic} in the equilibrium state~\cite{PhysRevD.50.6357}.

For the time evolution of the correlation functions, recurrence relations for $a_{n, k}$ are obtained by substituting the expansion Eq.~(\ref{eq:stochinf_expv}) into Eq.~(\ref{eq:stochinf_expvrec}). 
For the two potentials, we find
\begin{alignat}{2}
    (k + 1) a_{n, k+1} &= \displaystyle - \frac{m^2}{3 H^2} n a_{n, k} + \frac{H^2}{8 \pi^2} n (n-1) a_{n-2, k} & \qquad & \text{(quadratic)} \,\,, \\
    (k + 1) a_{n, k+1} &= \displaystyle - \frac{\lambda}{3 H^2} n a_{n+2, k} + \frac{H^2}{8 \pi^2} n (n-1) a_{n-2, k} & \qquad & \text{(quartic)} \,\,.
    \label{eq:stochinf_coefrec}
\end{alignat}
In the following, we will focus for definiteness on the two-point function, the power spectrum (and the corresponding $a_{2,k}$).
We will also apply the same tools, following the same steps, to the four-point function, the trispectrum (and the corresponding $a_{4,k}$).
In principle, the time dependence of correlation functions of any order $n$ can be studied by these means. 
\begin{table}
  \centering
  \begin{tabular}{|c||c|c|c|c|c|c|c|c} \hline 
  $n \backslash k$ & $0$ & $1$ & $2$ & $3$ & $4$ & $5$ & $6$ & $\cdots$ \\ \hline\hline
  $0$ & $\cellcolor{blue!25} 1$ & $\cellcolor{blue!25} 0$ & $\cellcolor{blue!25} 0$ & $\cellcolor{blue!25} 0$ & $\cellcolor{blue!25} 0$ & $\cellcolor{blue!25} 0$ & $\cellcolor{blue!25} 0$ & $\cellcolor{blue!25} \cdots$ \\ \hline
  $1$ & $\cellcolor{blue!25} 0$ & $\cellcolor{yellow!25} 0$ & $\cellcolor{yellow!25} 0$ & $\cellcolor{yellow!25} 0$ & $\cellcolor{yellow!25} 0$ & $\cellcolor{yellow!25} 0$ & $\cellcolor{yellow!25} 0$ & $\cellcolor{yellow!25} \cdots$ \\ \hline
  $2$ & $\cellcolor{blue!25} 0$ & $\cellcolor{red!25} 1 / 4 \pi^2$ & $- 1 / 12 \pi^2$ & $1 / 54 \pi^2$ & $- 1 / 324 \pi^2$ & $1 / 2430 \pi^2 $ & $- 1 / 21870 \pi^2$ & $\cdots$ \\ \hline
  $3$ & $\cellcolor{blue!25} 0$ & $\cellcolor{yellow!25} 0$ & $\cellcolor{yellow!25} 0$ & $\cellcolor{yellow!25} 0$ & $\cellcolor{yellow!25} 0$ & $\cellcolor{yellow!25} 0$ & $\cellcolor{yellow!25} 0$ & $\cellcolor{yellow!25} \cdots$ \\ \hline
  $4$ & $\cellcolor{blue!25} 0$ & $\cellcolor{yellow!25} 0$ & $\cellcolor{red!25} 3 / 16 \pi^4$ & $- 1 / 8 \pi^4$ & $7 / 144 \pi^4$ & $- 1 / 72 \pi^4$ & $31 / 9720 \pi^4$ & $\cdots$ \\ \hline
  $5$ & $\cellcolor{blue!25} 0$ & $\cellcolor{yellow!25} 0$ & $\cellcolor{yellow!25} 0$ & $\cellcolor{yellow!25} 0$ & $\cellcolor{yellow!25} 0$ & $\cellcolor{yellow!25} 0$ & $\cellcolor{yellow!25} 0$ & $\cellcolor{yellow!25} \cdots$ \\ \hline
  $6$ & $\cellcolor{blue!25} 0$ & $\cellcolor{yellow!25} 0$ & $\cellcolor{yellow!25} 0$ & $\cellcolor{red!25} 15 / 64 \pi^6$ & $-15 / 64 \pi^6$ & $25 / 192 \pi^6$ & $- 5 / 96 \pi^6$ & $\cdots$ \\ \hline
  $\vdots$ & $\cellcolor{blue!25} \vdots$ & $\cellcolor{yellow!25} \vdots$ & $\cellcolor{yellow!25} \vdots$ & $\cellcolor{yellow!25} \vdots$ & $\vdots$ & $\vdots$ & $\vdots$ & $\ddots$ 
  \end{tabular}
  \caption{
  The coefficients $\bar{a}_{n, k}$ for $V (\phi) = m^2 \phi^2 / 2$. 
  The colored entries are immediately determined from the initial and boundary conditions as well as the recurrence relations. 
  }
  \label{tbl:phi2}
  \end{table}

\begin{figure}
    \begin{tabular}{cc}
      \begin{minipage}[t]{0.495\hsize}
        \centering
        \subcaption{Truncation for $V (\phi) = m^2 \phi^2 / 2$.}
        \includegraphics[width = 0.88\linewidth]{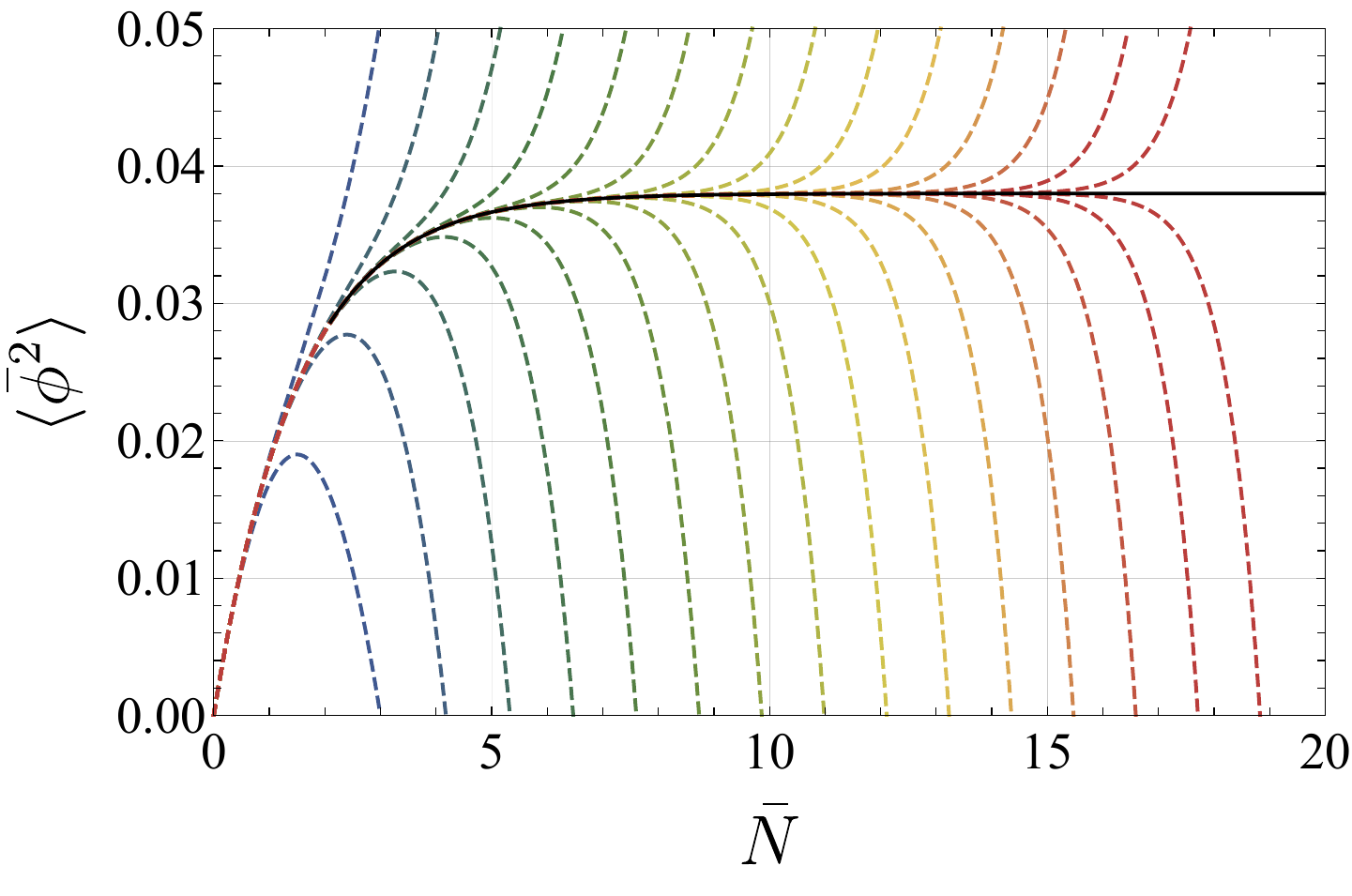}
      \end{minipage} &
      \begin{minipage}[t]{0.495\hsize}
        \centering
        \subcaption{Truncation for $V (\phi) = m^2 \phi^2 / 2$.}
        \includegraphics[width = 0.88\linewidth]{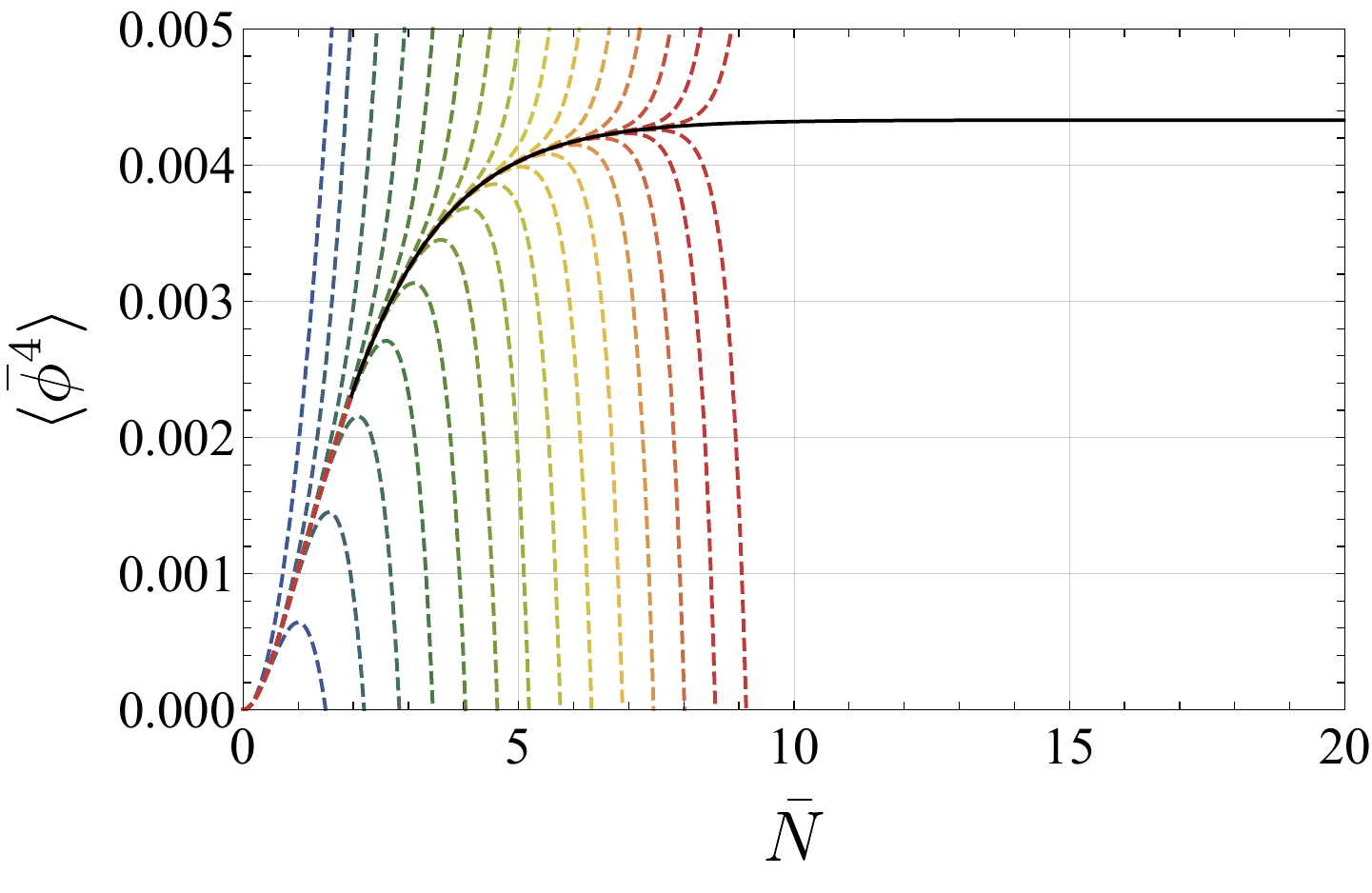}
      \end{minipage} \\
   
      \begin{minipage}[t]{0.495\hsize}
        \centering
        \subcaption{Truncation for $V (\phi) = \lambda \phi^4 / 4$.}
        \includegraphics[width = 0.88\linewidth]{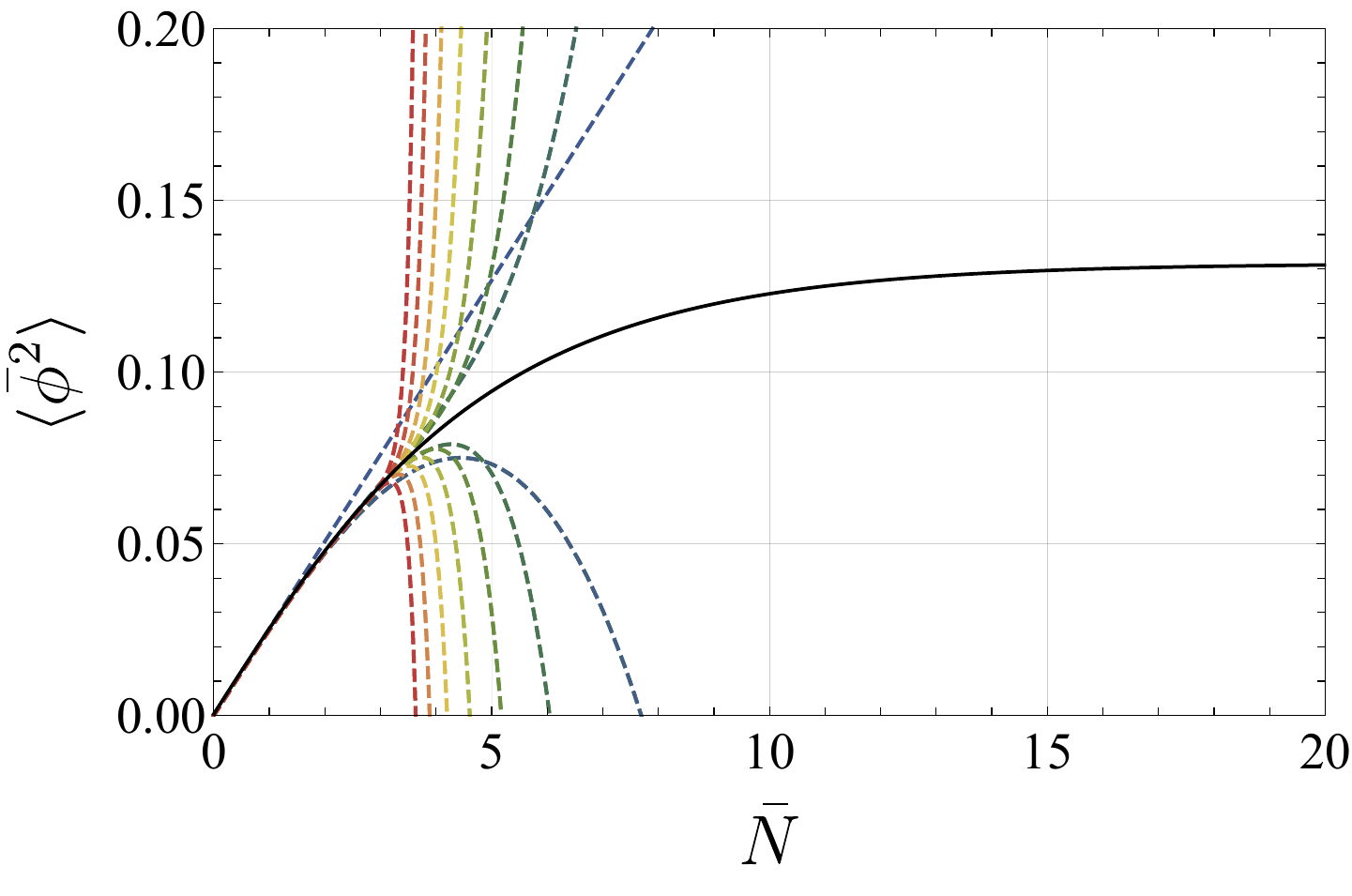}
      \end{minipage} &
      \begin{minipage}[t]{0.495\hsize}
        \centering
        \subcaption{Truncation for $V (\phi) = \lambda \phi^4 / 4$.}
        \includegraphics[width = 0.88\linewidth]{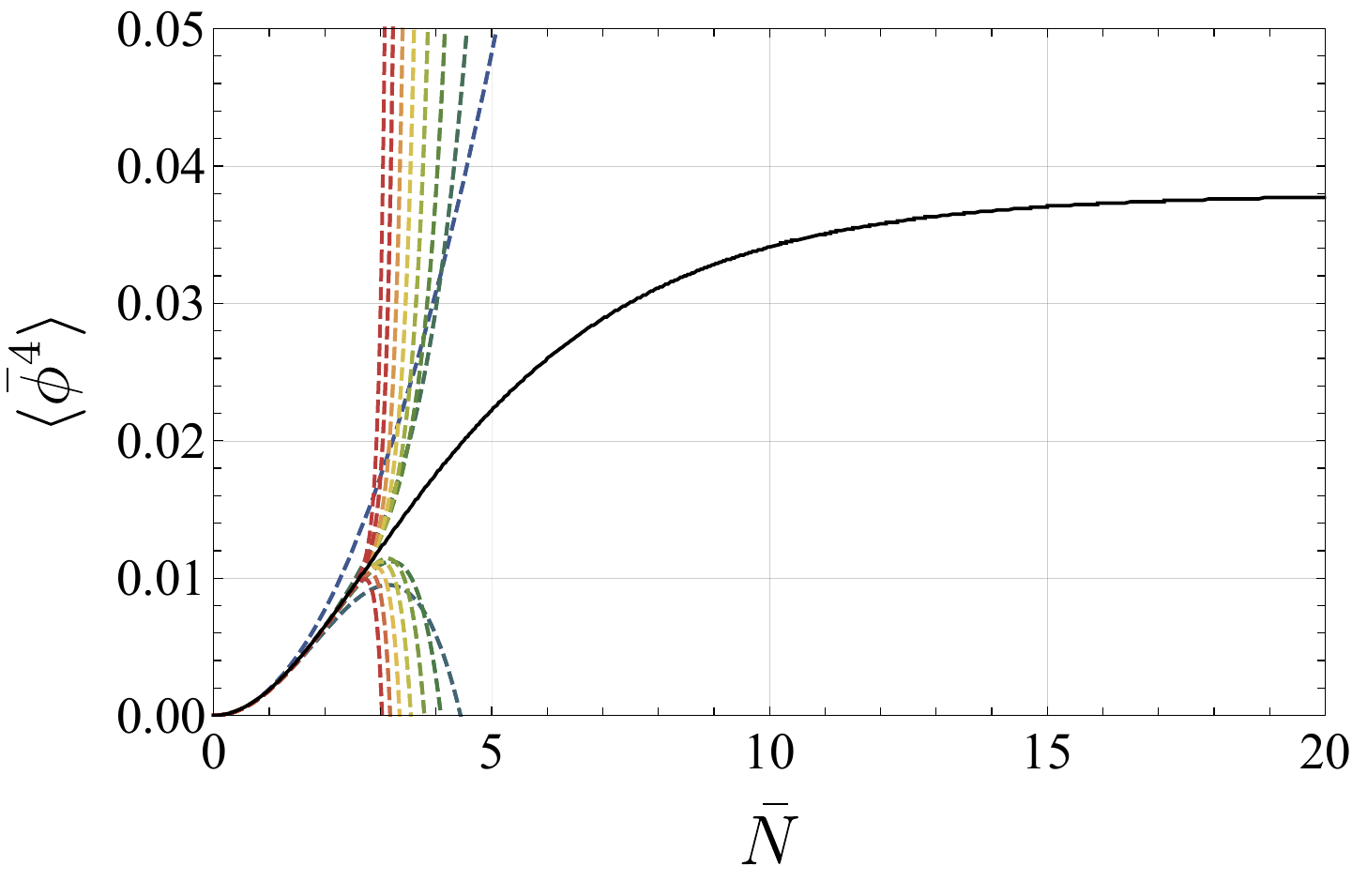}
      \end{minipage} 
    \end{tabular}
    \caption{
        Series truncation of the correlations $\expval{ \bar{\phi}^2 } (\bar{N})$ (\textit{left}) and $\expval{ \bar{\phi}^4 } (\bar{N})$ (\textit{right}), 
        for the quadratic (\textit{top}) and quartic (\textit{bottom}) potentials. 
        The truncation orders $k_{\rm T}$ are set to be $k_{\rm T} = 2, \, 3, \, \dots, \, 30$, and $k_{\rm T}$ increases from blue to red. 
    }
    \label{fig:truncation}
  \end{figure}

\paragraph{Quadratic case} 
For $V (\phi) = m^2 \phi^2 / 2$, the recurrence relation can be solved analytically to give
\begin{equation}
    \bar{a}_{2, k} 
    = \begin{cases}
        \displaystyle 0 & (k = 0) \\[2.0ex] 
        \displaystyle - \frac{3}{8 \pi^2} \frac{ ( - )^k }{k!} \qty( \frac{2}{3} )^k & (k \geq 1) 
    \end{cases}
     \,\, , \qquad 
     a_{n, k} = \bar{a}_{n, k} \frac{ m^{2k} }{ m^n } \frac{ H^{2n} }{H^{2k}} \,\, , 
     \label{eq:stochinf_tildequad}
\end{equation}
where we introduced the rescaled coefficients $\bar{a}_{n, k}$ so that the recurrence relation reduces to 
$(k + 1) \bar{a}_{n, k+1} = - (1/3) n \bar{a}_{n, k} + (1 / 8 \pi^2) n (n-1) \bar{a}_{n - 2, k}$ for $n \geq 2$ and $k \geq 0$.  
From Eq.~(\ref{eq:stochinf_tildequad}), the time dependence of the two-point function reads 
\begin{equation}
  \expval{ \phi^2 } (N) 
  = \sum_{k = 0}^{\infty} a_{2, k} N^k 
  = \frac{3 H^4}{8 \pi^2 m^2} \qty[ 1 - \exp \qty( - \frac{2 m^2}{3 H^2} N ) ] \,\, , 
  \label{eq:stochinf_expvquad}
\end{equation}
and $\expval{ \phi^4 } = 3 \expval{ \phi^2 }^2$. 
The same expression can be obtained without expanding $\expval{ \phi^2 }$ in terms of $N$, by directly solving Eq.~(\ref{eq:stochinf_expvrec}), the last term in the right hand side being a constant for $n = 2$. 
Our conclusion regarding the expansion of $\expval{ \phi^2 }$ in terms of $N$, is that it results in a convergent series as it should be, therefore giving the exact formula Eq.~(\ref{eq:stochinf_expvquad}).
Any higher order correlation function can be computed this way for the quadratic case (another option is to compute them  from the Gaussian density function $f(\phi,N)$), see Table~\ref{tbl:phi2} for a few other values of the $\bar{a}_{n,k}$.
\begin{figure}
  \begin{minipage}[b]{0.495\linewidth}
    \centering
    \subcaption{
        Coefs. for $V (\phi) = m^2 \phi^2 / 2$.
    }
    \includegraphics[width = 0.95\linewidth]{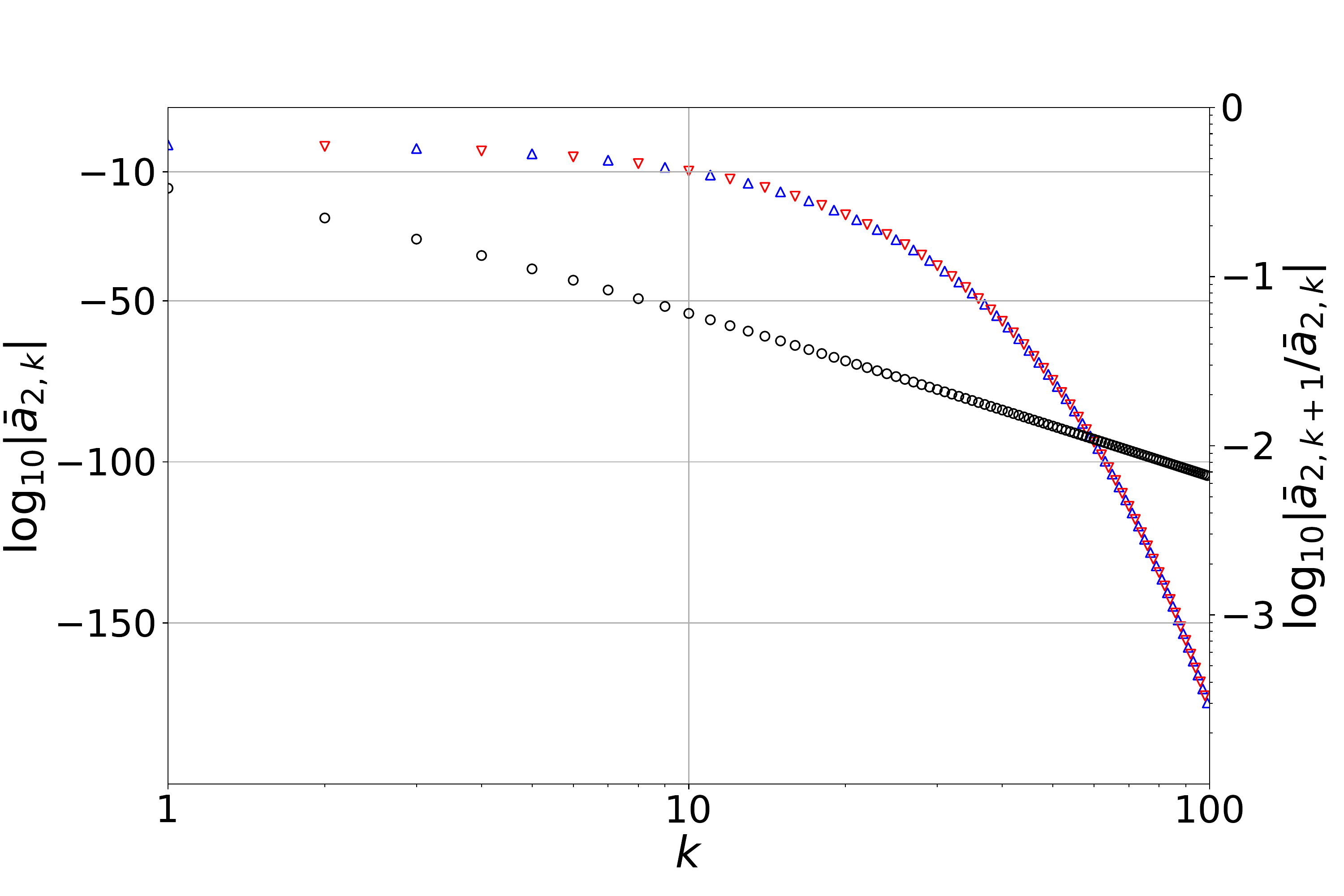}
  \end{minipage}
  \begin{minipage}[b]{0.495\linewidth}
    \centering
    \subcaption{
        Coefs. for $V (\phi) = \lambda \phi^4 / 4$.
    }
    \includegraphics[width = 0.95\linewidth]{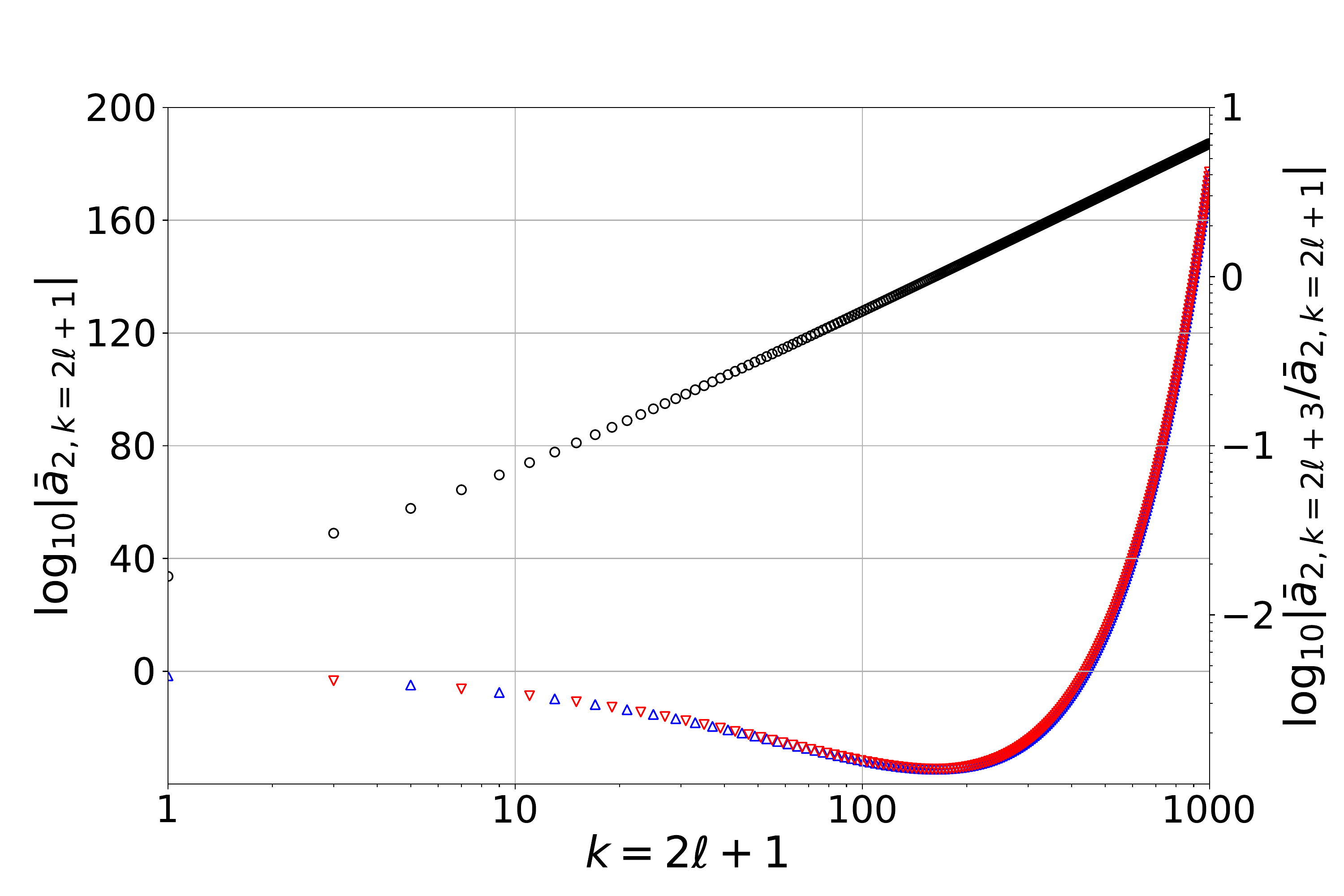}
  \end{minipage}
  \caption{
    The coefficients $\bar{a}_{2, k}$ and their ratio to the previous order, for the quadratic (\textit{left}) and quartic (\textit{right}) potentials. 
    The upward and downward triangles indicate the sign of $\bar{a}_{2, k}$, and their absolute values can be read off from the left vertical axis.
    The circles are the ratio between neighboring two nonzero coefficients and can be read off from the right vertical axis. 
  }
  \label{fig:coefficients_p2}
\end{figure}
The left panel of Fig.~\ref{fig:coefficients_p2} shows the behaviour of the coefficients $\bar{a}_{2,k}$ and their ratio $\bar{a}_{2,k+1} / \bar{a}_{2,k}$ as $k$ increases.
Noting that the ratio is related to the convergence radius $r_c$ of the series by
\begin{equation}
\frac{1}{r_c} 
= \lim_{k \to \infty} \abs{ \frac{\bar{a}_{2,k+1}}{\bar{a}_{2,k}} } \,\, , 
\end{equation}
the plot implies that the expansion (\ref{eq:stochinf_expvquad}) indeed has an infinite convergence radius.
The same holds for the equivalent expressions for $\expval{ \phi^4 }$. 
For convenience, we introduce dimensionless variables,
\begin{equation}
  \expval{ \bar{ \phi }^2 } = \sum_{k = 0}^{\infty} \bar{a}_{2, k} \bar{N}^k \,\, ; 
  \qquad \bar{\phi} \equiv \frac{m}{H^2} \phi \,\, , 
  \qquad 
  \bar{N} \equiv \frac{m^2}{H^2} N \,\, , 
  \label{eq:stochinf_dimless2}
\end{equation}
then $a_{2, k} / a_{2, k + 1} = (H^2/m^2) \, \bar{a}_{2, k} / \bar{a}_{2, k + 1}$. 
The top panels of Fig.~\ref{fig:truncation} compares
the series truncation of $\expval{ \bar{\phi}^2 } (\bar{N})$ and $\expval{ \bar{\phi}^4 } (\bar{N})$ at different orders with their exact results. 
We easily see that the series truncated at higher orders give better approximations over the whole region, as expected from the fact that the series (\ref{eq:stochinf_dimless2}) has an infinite radius of convergence.\footnote{
In contrast, if the series has a finite radius of convergence, the expected behavior is that higher order truncations approximate the exact result more accurately for $\bar{N}$ smaller than the convergence radius, and then they deviate from the exact result with a blowup for larger $\bar{N}$.
}
\begin{table}
  \centering
  \begin{tabular}{|c||c|c|c|c|c|c|c|c} \hline 
  $n \backslash k$ & $0$ & $1$ & $2$ & $3$ & $4$ & $5$ & $6$ & $\cdots$ \\ \hline\hline
  $0$ & $\cellcolor{blue!25} 1$ & $\cellcolor{blue!25} 0$ & $\cellcolor{blue!25} 0$ & $\cellcolor{blue!25} 0$ & $\cellcolor{blue!25} 0$ & $\cellcolor{blue!25} 0$ & $\cellcolor{blue!25} 0$ & $\cellcolor{blue!25} \cdots$ \\ \hline
  $1$ & $\cellcolor{blue!25} 0$ & $\cellcolor{yellow!25} 0$ & $\cellcolor{yellow!25} 0$ & $\cellcolor{yellow!25} 0$ & $\cellcolor{yellow!25} 0$ & $\cellcolor{yellow!25} 0$ & $\cellcolor{yellow!25} 0$  & $\cellcolor{yellow!25} \cdots$ \\ \hline
  $2$ & $\cellcolor{blue!25} 0$ & $\cellcolor{red!25} 1 / 4 \pi^2$ & $0$ & $- 1 / 24 \pi^4$ & $0$ & $1 / 80 \pi^6$ & $0$ & $\cdots$ \\ \hline
  $3$ & $\cellcolor{blue!25} 0$ & $\cellcolor{yellow!25} 0$ & $\cellcolor{yellow!25} 0$ & $\cellcolor{yellow!25} 0$ & $\cellcolor{yellow!25} 0$ & $\cellcolor{yellow!25} 0$ & $\cellcolor{yellow!25} 0$ & $\cellcolor{yellow!25} \cdots$ \\ \hline
  $4$ & $\cellcolor{blue!25} 0$ & $\cellcolor{yellow!25} 0$ & $\cellcolor{red!25} 3 / 16 \pi^4$ & $0$ & $- 3 / 32 \pi^6$ & $0$ & $53 / 960 \pi^8$ & $\cdots$ \\ \hline
  $5$ & $\cellcolor{blue!25} 0$ & $\cellcolor{yellow!25} 0$ & $\cellcolor{yellow!25} 0$ & $\cellcolor{yellow!25} 0$ & $\cellcolor{yellow!25} 0$ & $\cellcolor{yellow!25} 0$ & $\cellcolor{yellow!25} 0$ & $\cellcolor{yellow!25} \cdots$ \\ \hline
  $6$ & $\cellcolor{blue!25} 0$ & $\cellcolor{yellow!25} 0$ & $\cellcolor{yellow!25} 0$ & $\cellcolor{red!25} 15 / 64 \pi^6$ & $0$ & $- 15 / 64 \pi^8$ & $0$ & $\cdots$ \\ \hline
  $\vdots$ & $\cellcolor{blue!25} \vdots$ & $\cellcolor{yellow!25} \vdots$ & $\cellcolor{yellow!25} \vdots$ & $\cellcolor{yellow!25} \vdots$ & $\vdots$ & $\vdots$ & $\vdots$ & $\ddots$
  \end{tabular}
  \caption{
  The coefficients $\bar{a}_{n, k}$ for $V (\phi) = \lambda \phi^4 / 4$. 
  The colored entries are immediately determined from the initial and boundary conditions as well as the recurrence relations. 
  }
  \label{tbl:phi4}
\end{table}

\paragraph{Quartic case} 
For $V (\phi) = \lambda \phi^4 / 4$, the coefficients of $N^k$ can be rescaled as $a_{n, k} = \bar{a}_{n, k} H^n \lambda^{(2 k - n)/4}$, and the recurrence relation reduces to $(k + 1) \bar{a}_{n, k+1} = - (n/3) \bar{a}_{n + 2, k} + (1 / 8 \pi^2) n (n-1) \bar{a}_{n-2, k}$. 
The closed form for general $\bar{a}_{2, k}$ is too complicated for practical use but again can be obtained. 
For even indices $k = 0, \, 2, \, \cdots$ the coefficient $\bar{a}_{2, k}$ vanishes, while for odd $k = 1, \, 3, \, \cdots$ it starts with $\bar{a}_{2, k = 1} = 1$ and 
\begin{equation}
    \bar{a}_{2, k} 
    = \qty( \frac{3}{2 \pi^2} )^{1/2} 
    \frac{ (-)^{(k-1)/2} }{k!} 
    \qty( \frac{1}{24 \pi^2} )^{k/2} 
    \prod_{j = 0}^{\frac{k-3}{2}} \sum_{p_j = j + 2}^{p_{j + 1}} (2 p_{j}  - 2 j - 2) (2 p_{j}  - 2 j - 1) (2 p_{j}  - 2 j) \,\, ,
    \label{eq:stochinf_anarec}
\end{equation}
for $k = 3, \, 5 \, \cdots$.\footnote{
Explicitly written, the product is
\begin{align}
\prod_{j = 0}^{\frac{k-3}{2}} \sum_{p_j = j + 2}^{p_{j + 1}} (\,\cdots)
&=
\sum_{p_{(k - 3)/2} = (k + 1)/2}^{p_{(k - 1)/2}} (\,\cdots)
~\cdots~
\sum_{p_2 = 4}^{p_3} (\,\cdots)
\sum_{p_1 = 3}^{p_2} (\,\cdots)
\sum_{p_0 = 2}^{p_1} (\,\cdots) \,\, ,
\end{align}
where $p_{(k - 1)/2} = (k + 1)/2$.
}
In this expression, $p_{(k - 1)/2}$ is given by $p_{(k - 1)/2} = (k + 1)/2$. 
This closed form is obtained for the first time to the best of the authors' knowledge.
Table~\ref{tbl:phi4} shows the first few terms of $\bar{a}_{n, k}$ for the quartic case. 
The coefficients for $n=2$ can be obtained from Eq.~(\ref{eq:stochinf_anarec}) or iteratively from Eq.~(\ref{eq:stochinf_coefrec}), and the first terms are in precise agreement with the result from more detailed field theoretical calculations (see \cite{TSAMIS2005295, PhysRevD.79.044007} for the same recursive calculations and \cite{PhysRevD.76.043512} for field theoretical derivations). 
Although one may expect from Table~\ref{tbl:phi4} that the $e$-folding expansion of $\expval{ \phi^2 }$ is again convergent, it is not the case. 
In order to see this, we introduce dimensionless variables similar to Eq.~(\ref{eq:stochinf_dimless2}), 
\begin{equation}
    \expval{ \bar{ \phi }^2 } = \sum_{k = 0}^{\infty} \bar{a}_{2, k} \bar{N}^k \,\, ; 
    \qquad 
    \bar{ \phi } \equiv \frac{ \lambda^{1/4} }{H} \phi \,\, , 
    \qquad 
    \bar{ N } \equiv \lambda^{1/2} N \,\, . 
    \label{eq:stochinf_expvquar}
\end{equation}
The right panel of Fig.~\ref{fig:coefficients_p2} shows $\bar{a}_{2, k = 2\ell + 1}$ and $\bar{a}_{2, k = 2\ell + 3} / \bar{a}_{2, k = 2\ell + 1}$ for the case of the quartic potential.
We see that the ratio exhibits a power-law growth with a positive exponent, which is typical of factorially divergent series that appear in various systems in physics.
Therefore the plot implies that the convergent radius of the expansion Eq.~(\ref{eq:stochinf_expvquar}) is zero in contrast to the quadratic case. 
Behaviour consistent with this can be seen in the bottom panels of Fig.~\ref{fig:truncation}, which compare truncated series with the numerical results. 
We see that higher order truncations start to blow up for smaller $\bar{N}$.
This is typical behaviour of a series with a vanishing radius of convergence, and its naive summation to the infinite order does not make sense; 
we just get infinity everywhere except $\bar{N}=0$. 
This calls for some resummation prescription in order to recover the correct time evolution.
\section{Time evolution of the correlation function from resummation}
\label{sec:section3}

The closed form of $\bar{a}_{2, k}$ obtained in Eq.~(\ref{eq:stochinf_anarec}) for the quartic potential gives the information on all orders, and thus in principle gives the \textit{exact} correlation function $\expval{\bar{\phi}^n}$. 
As mentioned in Sec.~\ref{sec:section2}, however, it is practically difficult to obtain all the coefficients analytically, and so is the time evolution of the correlation function.
To make matters worse, its $e$-folding expansion is a formal perturbative series that deviates more and more from its original behaviour as the order of truncation increases, as shown in the bottom panels of Fig.~\ref{fig:truncation}.
In order to tame the divergence, we consider two kinds of resummation methods in this section, namely the Pad\'{e} approximant and the Borel--Pad\'{e} resummation.

\subsection{\textit{Pad\'{e} approximant}: Approximating the exact behaviour by a rational function}

The \textit{Pad\'{e} approximant}~\cite{baker1996pade} approximates a function by a rational function, in such a way that its power series agrees with that of the original function up to a given order.
It often gives a better approximation of the original function than the na\"{i}vely truncated series, and it can be used even when the power series of the original function is divergent. 
We apply this procedure to the correlation functions of the spectator field.

The Pad\'{e} approximant is constructed as follows.
With a pair of integers $m, \, n \in \mathbb{Z}_{\geq 0}$, a smooth function $f (z)$ is approximated by a rational function called the Pad\'{e} approximant,
\begin{equation}
  f (z) = \sum_{k = 0}^{\infty} a_k z^k 
  \quad 
  \longrightarrow 
  \quad 
  f^{[m|n]}_{\mathrm{P}} (z) 
  \equiv 
  \frac{ \displaystyle \sum_{k = 0}^{m} b_k z^k }{ \displaystyle \sum_{k = 0}^{n} c_k z_k }
  = \frac{ b_0 + b_1 z + b_2 z^2 + \cdots + b_m z^m }{c_0 + c_1 z + c_2 z^2 + \cdots + c_n z^n } \,\, . 
  \label{eq:Pade_def}
\end{equation}
We require that $f (z)$ and $f_{\mathrm{P}}^{[m | n]} (z)$ be related by $f (z) = f_{\mathrm{P}}^{[m | n]} (z) + \mathcal{O} (z^{m + n + 1})$. 
Thus we need the first $m + n + 1$ coefficients of the Taylor expansion of $f (z)$ around $z=0$, and the coefficients $b_k$ and $c_k$ in Eq.~(\ref{eq:Pade_def}) are uniquely determined by\footnote{
Here we have implicitly assumed that $f(z)$ is Taylor-expandable around $z=0$, and this holds true for the problems studied in this paper.
If this is not the case, we could not directly use the Pad\'{e} approximant.
A typical case is when the asymptotic behaviour of $f(z)$ around $z = 0$ is singular, {\it e.g.}~$f(z) = \sqrt{z}$, $1/z$, or $e^{1/z}$.
Even in such cases, the problem can often be reduced to an equivalent one such that the Pad\'{e} approximant is applicable by an appropriate mapping.
For instance, when $f(z) = \sqrt{z} g(z)$ with $g(z)$ Taylor-expandable around $z = 0$, we can apply the Pad\'{e} approximant for $g(z) = f(z)/\sqrt{z}$. 
}
\begin{equation}
  \eval{ \dv[\ell]{ f (z) }{z} }_{z = 0} 
  = \eval{ \dv[\ell]{ f_{\mathrm{P}}^{[m|n]} (z) }{z} }_{z = 0} \,\, , 
  \qquad 
  \ell = 0, \, 1, \, \dots, \, m + n \,\, .
  \label{eq:Pade_coef}
\end{equation}
The set of conditions (\ref{eq:Pade_coef}) can be translated into 
\begin{equation}
    \sum_{k = 0}^{\ell} a_k c_{\ell - k} - b_{\ell} = 0 \,\, , 
    \qquad 
    \ell = 0, \, 1, \, \dots, \, m + n \,\, . 
    \label{eq:Pade_coefd}
\end{equation}
It is clear that $f^{[m|n]}_{\mathrm{P}} (z)$ reduces to the $m$-th order of the Taylor expansion of $f (z)$ when $n = 0$. 
The special case, $m = n$, is called the \textit{diagonal} Pad\'{e} approximant, and it is known to often give a better approximation than the ones with $m \neq n$ called non-diagonal Pad\'{e} approximants.
In the following, however, we restrict ourselves to the non-diagonal choice with $m = p - 1$ and $n = p + 1$ for a positive integer $p$. 
This choice is made for the purpose of using the same order for the Pad\'{e} approximants consistently throughout the paper: since the order of the Pad\'{e} approximants used in Borel--Pad\'{e} resummation, as we see in Sec.~\ref{sec:Borel}, is required to satisfy $m < n$ from the viewpoint of convergence, we use the same choice here.
For completeness, in Appendix~\ref{app:appendix1} we also show the result of diagonal Pad\'{e} for $\expval{\phi^2}$ and $\expval{\phi^4}$.

Pad\'{e} approximants have several important properties essentially coming from the fact that they are rational functions.
First, Pad\'{e} approximants cannot have branch cuts while they can have poles.
This implies that when we try to approximate a function with branch cuts, Pad\'{e} approximants cannot reproduce exactly the same analytic structure as the original function has.
Instead, higher-order Pad\'{e} approximants typically develop a bunch of poles and/or zeros around the location of the branch cut of the original function.\footnote{
See \textit{e.g.} \cite{yamada2014numerical} for some benchmarks.
}
For this reason, Pad\'{e} approximants typically give better approximations for meromorphic functions than for functions with branch cuts.
Second, the series expansion of a Pad\'{e} approximant with a finite order around the origin is always convergent.
This means that, when the original function has a divergent series around the origin, Pad\'{e} approximants with a finite order cannot share this property.
Thus, if we know some of the properties of the original function {\it a priori}, it is better to adopt an approximation scheme that correctly captures these properties.\footnote{
There are various approximation schemes beyond the standard Pad\'{e} approximation, see \cite{Sen:2013oza,Beem:2013hha,Honda:2014bza,Honda:2015ewa,Alday:2013bha,Chowdhury:2016hny,Costin:2020hwg,Costin:2020pcj,Costin:2021bay,Costin:2022hgc}.
}
If otherwise, 
Pad\'{e} approximants are usually a good first step to probe some of the properties. 

\begin{figure}
    \begin{tabular}{cc}
      \begin{minipage}[t]{0.495\hsize}
        \centering
        \subcaption{Direct Pad\'{e} for $V (\phi) = m^2 \phi^2 / 2$.}
        \includegraphics[width=0.88\linewidth]{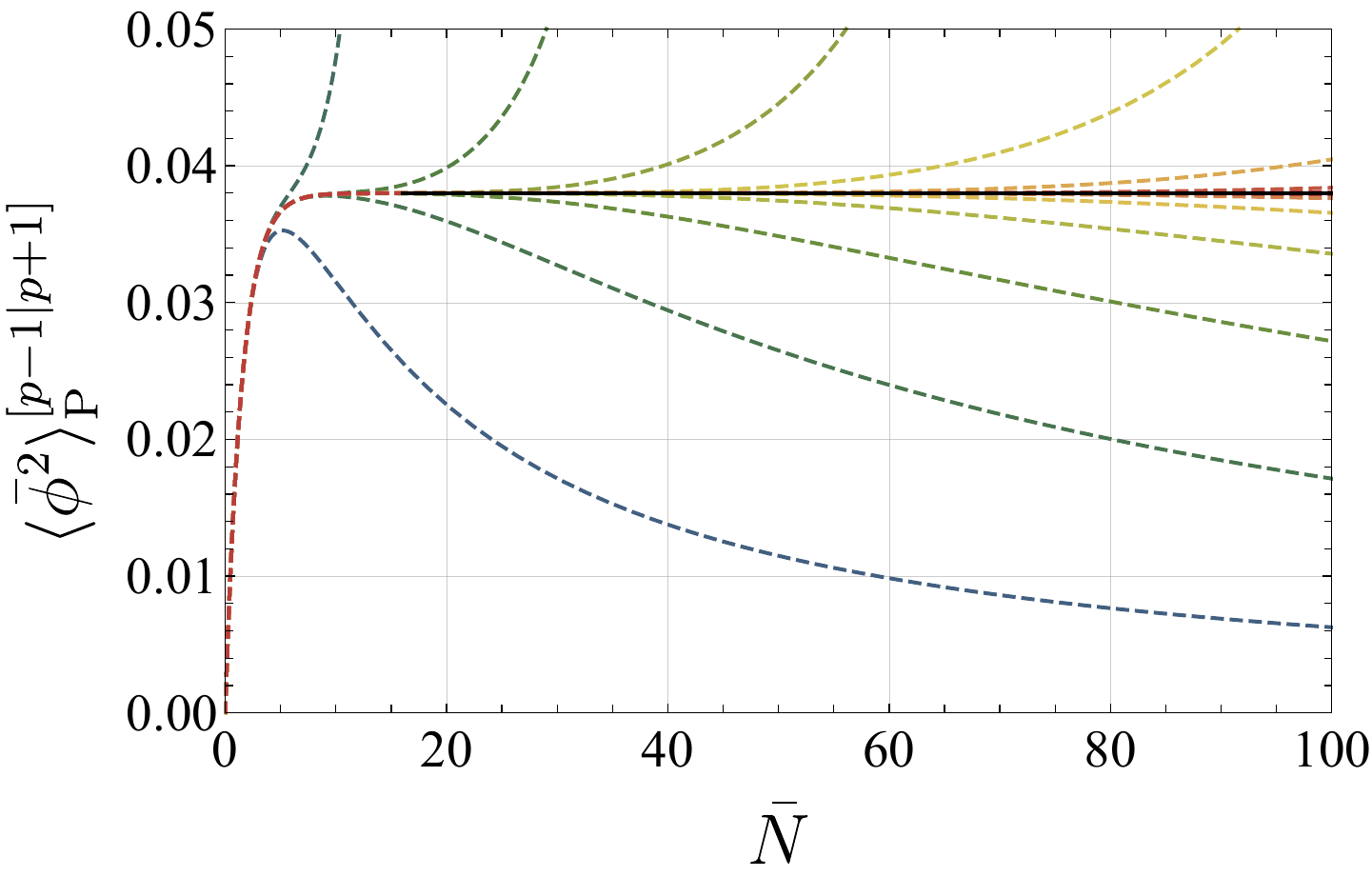}
      \end{minipage} &
      \begin{minipage}[t]{0.495\hsize}
        \centering
        \subcaption{Direct Pad\'{e} for $V (\phi) = m^2 \phi^2 / 2$.}
        \includegraphics[width=0.88\linewidth]{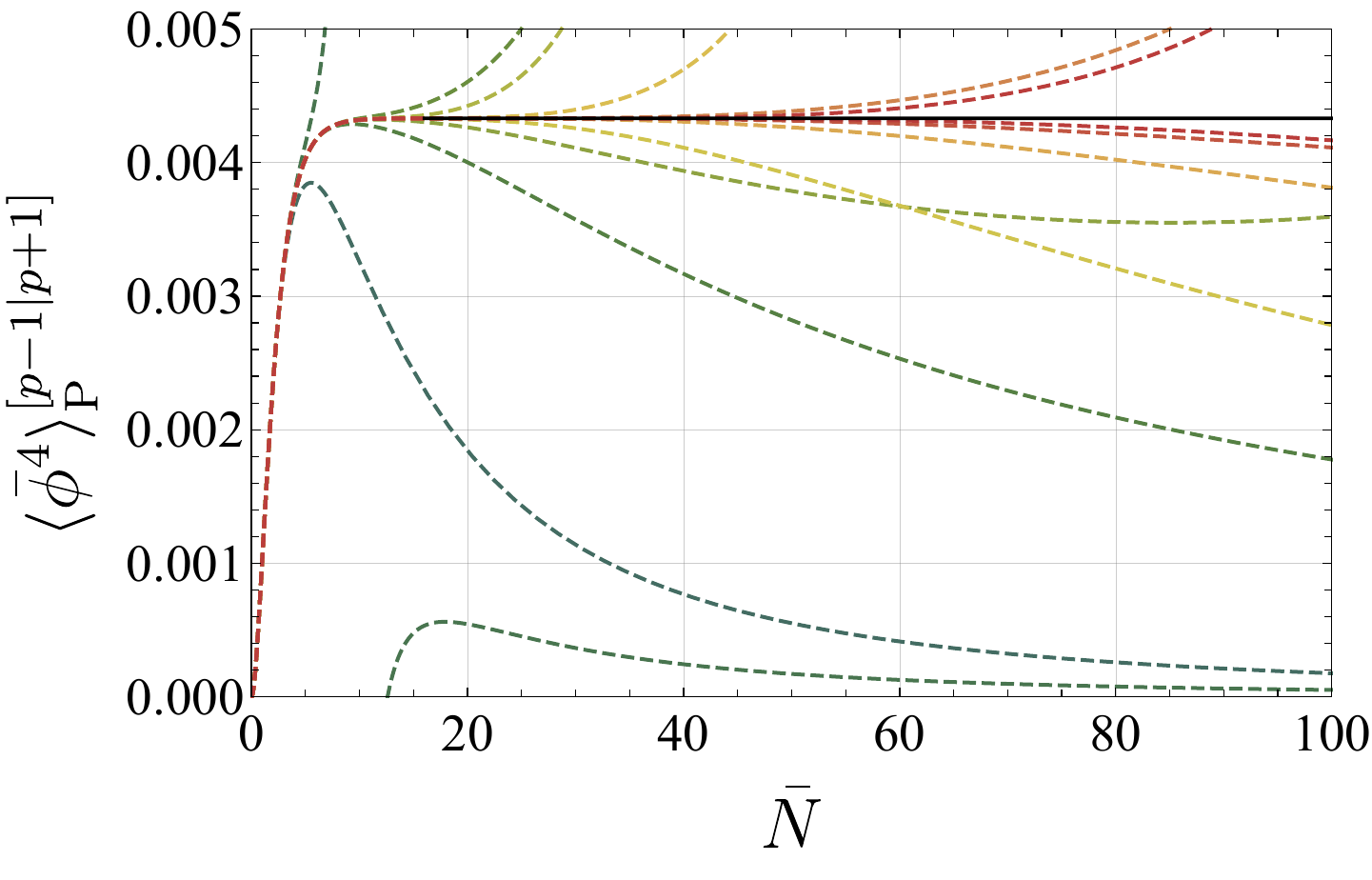}
      \end{minipage} \\
      \begin{minipage}[t]{0.495\hsize}
        \centering
        \subcaption{Direct Pad\'{e} for $V (\phi) = \lambda \phi^4 / 4$.}
        \includegraphics[width = 0.88\linewidth]{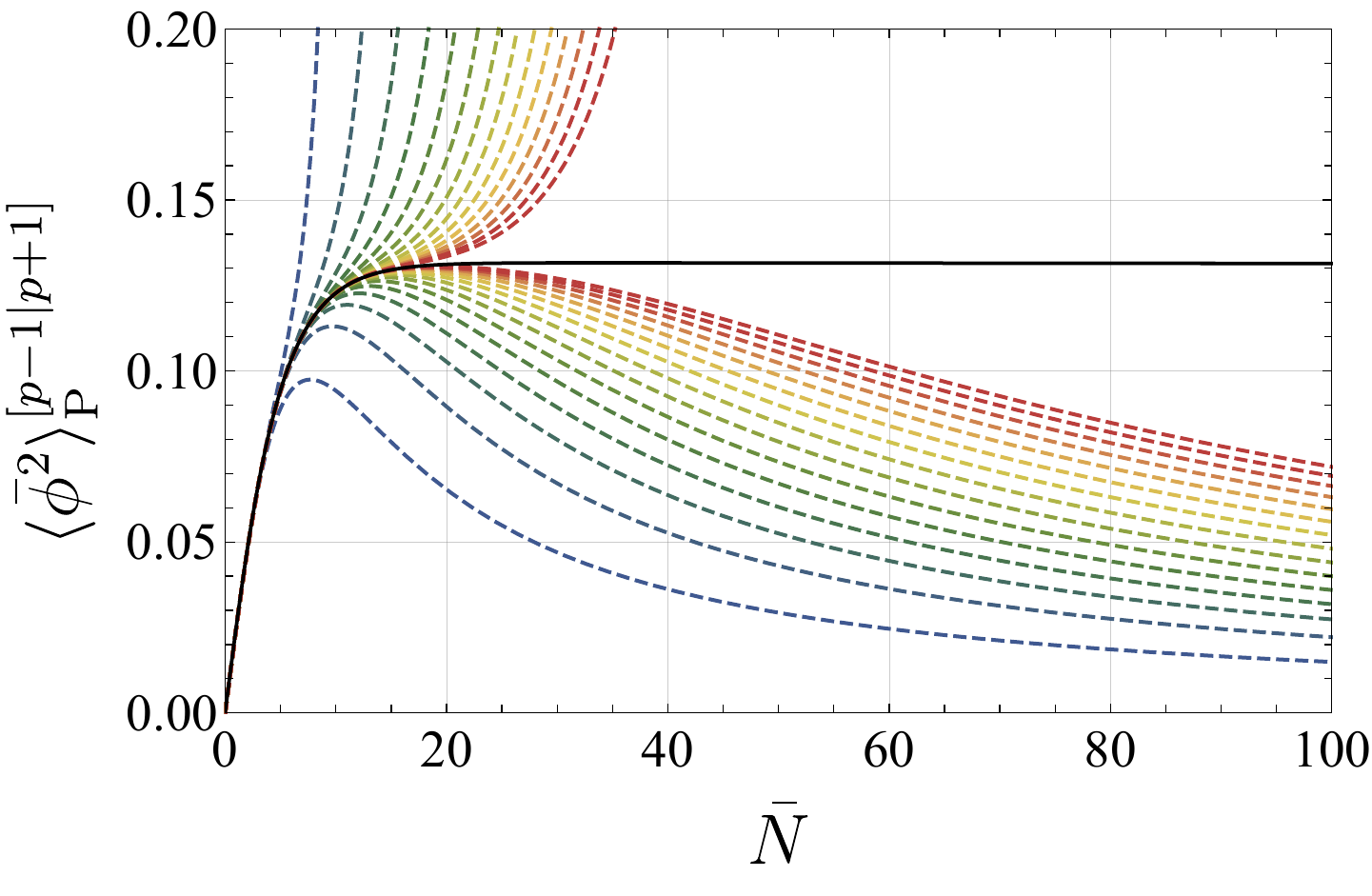}
      \end{minipage} &
      \begin{minipage}[t]{0.495\hsize}
        \centering
        \subcaption{Direct Pad\'{e} for $V (\phi) = \lambda \phi^4 / 4$.}
        \includegraphics[width = 0.88\linewidth]{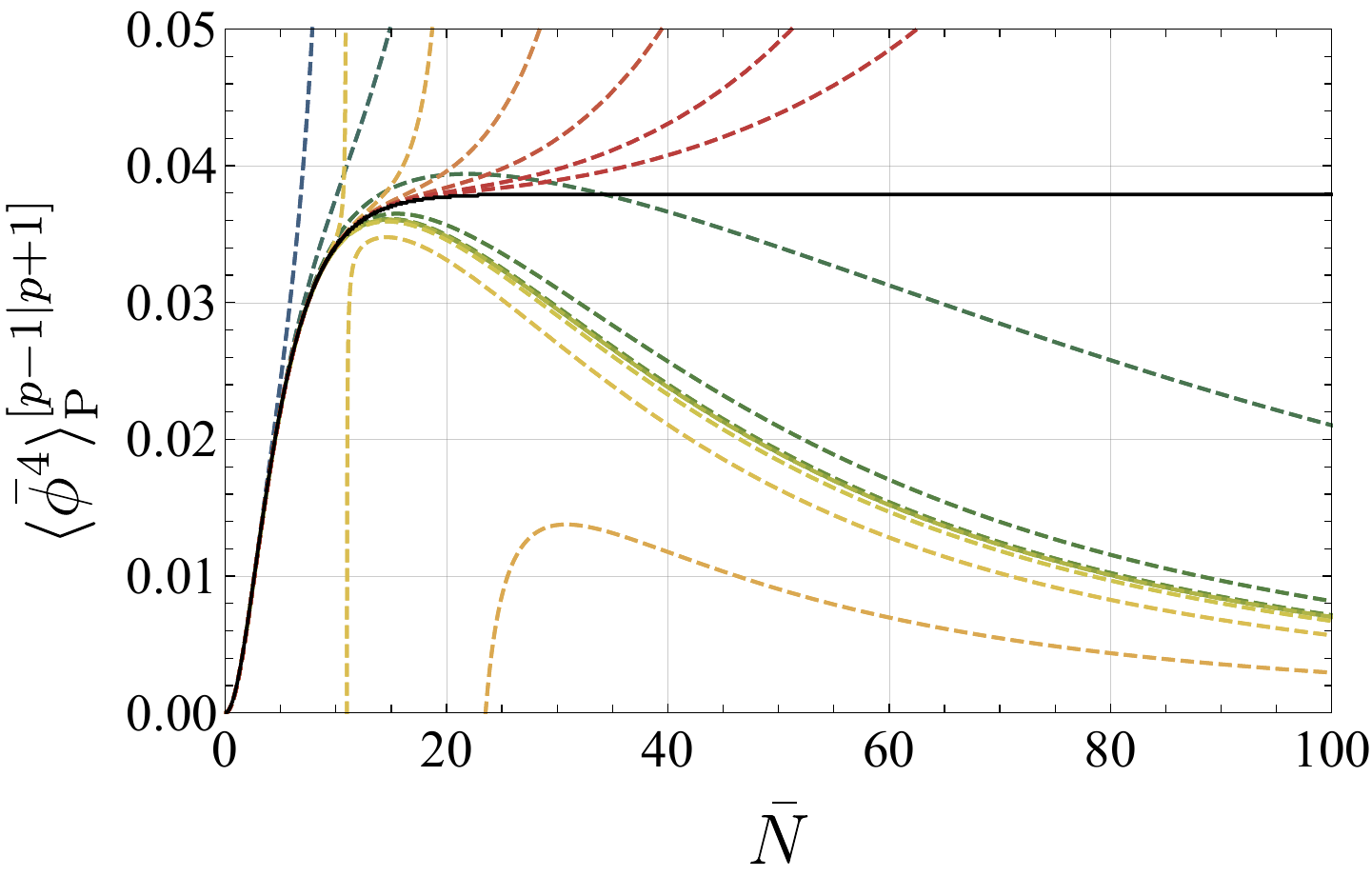}
      \end{minipage} 
    \end{tabular}
     \caption{
        \textit{Direct} Pad\'{e} approximants $\expval{\bar{\phi}^2}_{\textrm{P}}^{[p - 1 | p + 1]} (\bar{N})$ (\textit{left}) and $\expval{\bar{\phi}^4}_{\textrm{P}}^{[p - 1 | p + 1]} (\bar{N})$ (\textit{right}), for the quadratic (\textit{top}, up to $p = 15$) and quartic (\textit{bottom}, up to $p = 30$) potentials. 
        The order $p$ increases from blue to red. 
     }
     \label{fig:DirectPade}
\end{figure}

From the above considerations, we construct the Pad\'{e} approximants of the correlation functions as
\begin{equation}
  \expval{ \phi^2 } (N) 
  = \sum_{k = 0}^{\infty} a_{2, k} N^k 
  \quad 
  \longrightarrow 
  \quad 
  \expval{ \phi^2 }_{\mathrm{P}}^{[p-1 | p+1]} (N)
  = \frac{ \displaystyle \sum_{k = 0}^{p-1} b_k N^k }{ \displaystyle \sum_{k = 0}^{p+1} c_k N^k } \,\, ,
\end{equation}
and similarly for $\expval{\phi^4}$.
Figure~\ref{fig:DirectPade} shows the Pad\'{e} approximant for both the quadratic and quartic potentials. 
Compared to Fig.~\ref{fig:truncation}, those indeed give improved behaviour compared to the na\"{i}vely truncated cases. 
Not only does the Pad\'{e} approximant reproduce the transient regime around $\bar{N} \sim 10$, but it also gives the correct stationary behaviour, especially for the quadratic case. 
However, quantitatively, we see a difference in accuracy between the quadratic and quartic cases.
In the quartic case, the approximation is relatively worse despite it uses higher order information, though it is still much better than the truncated series.
In the quadratic case, the exact results are entire functions and the Pad\'{e} approximant is good at approximating such functions.
As mentioned above, the quartic case has a divergent perturbative series and its Pad\'{e} approximant with a finite order cannot have such a property.
Therefore the Pad\'{e} approximant is likely worse at approximating functions having divergent series compared to analytic functions.
This motivates us to consider another resummation scheme that efficiently takes the properties of the series into account.
One may wonder what if we use diagonal Pad\'{e}, since in the diagonal case the highest orders of the numerator and denominator in Eq.~(\ref{eq:Pade_def}) are the same, and thus the asymptotic stationary behavior of the correlation functions is guaranteed.
Interestingly, however, this improvement applies only to some of the correlation functions (more specifically, $\expval{\phi^4}$, $\expval{\phi^8}$, $\cdots$).
We illustrate this point in Appendix~\ref{app:appendix1}.

\subsection{\textit{Borel resummation}: Extracting the correct information from a formal series}
\label{sec:Borel}

In the last subsection we saw how the Pad\'{e} approximant reproduces the original correlation functions up to some moderate $e$-folding number, even when the original power series is divergent and defined only formally.
However, we also found that in the quartic case the approximation is relatively worse presumably because this case has a divergent perturbative series and the Pad\'{e} approximant is likely worse at approximating such functions than analytic functions.  
Here we take another strategy: Borel--Pad\'{e} resummation.
It is a combination of Borel resummation and Pad\'{e} approximation, and there are also other motivations beyond the above to use it in the present context. 
Here, we take another strategy: \textit{Borel resummation} (or practically Borel--Pad\'{e} resummation, use of Pad\'{e} approximation in Borel summation.) 
There are also other motivations beyond the above to use it in the present context.
First, correlation functions in de Sitter spacetime are expected to reach asymptotic values for sufficiently large $N$.
As we see below, the Borel transformation of such functions in general converges to zero in the Borel plane.
This property makes it easier to approximate the function with the Pad\'{e} approximation, and thus we expect that the behaviour of the divergent series is improved even more with Borel--Pad\'{e} resummation.
Second, Borel transformation contains information about possible non-perturbative aspects of the system through the singularity structure in the Borel plane, and thus is physically interesting to investigate.
In the following we illustrate how this method works for the correlation functions in de Sitter space.

Borel resummation~\cite{ASENS_1899_3_16__9_0, bender78:AMM, SPT:KT} is defined through the Borel transformation of the original series.\footnote{
    It is convenient to put $\alpha \notin \mathbb{Z}_{\leq 0}$ to the index of $z$, following the definition of \cite{SPT:KT}.
}
For an infinite series with respect to $z$ and $\alpha \notin \mathbb{Z}_{\leq 0}$, 
\begin{equation}
  f (z) = \sum_{k = 0}^{\infty} a_k z^{k + \alpha} \,\, , 
  \label{eq:Borel_series}
\end{equation}
we define the \textit{Borel transformation} of $f (z)$ by 
\begin{equation}
  f_{\mathrm{B}} (t) 
  \equiv \sum_{k = 0}^{\infty} \frac{a_{k}}{\Gamma (k + \alpha)} t^{k + \alpha - 1} 
  \,\, . 
  \label{eq:Borel_defBT}
\end{equation}
Then the \textit{Borel resummation} of $f (z)$ is defined as
\begin{equation}
  f_{\mathrm{S}} (z) 
  \equiv \int_{0}^{\infty} \dd t \, e^{- t / z} \widetilde{f_{\mathrm{B}}} (t) \,\, , 
  \label{eq:Borel_defBS}
\end{equation}
where $\widetilde{f_{\rm B}} (t)$ is a simple analytic continuation of the series (\ref{eq:Borel_defBT}).
The Borel resummation (\ref{eq:Borel_defBS}) has the following important properties.
First, it has the same asymptotic behavior around $z=0$ as the original one (\ref{eq:Borel_series}) (up to exponentially suppressed corrections that may appear).
One can easily check this by expanding $\widetilde{f_{\rm B}} (t)$ around $t=0$ as in Eq.~\eqref{eq:Borel_defBT} and then exchanging the order of the $t$-integration and the expansion.
Second, the Borel resummation can be finite for finite and non-zero $z$ under some conditions (explained later), even if the original perturbative series (\ref{eq:Borel_series}) is divergent.
Because of these reasons, the Borel resummation may correctly capture the true properties of the original function and has turned out to be the most standard way to resum divergent perturbative series.

Let us emphasize the contexts in which Borel resummation works.
The Borel resummation, Eq.~(\ref{eq:Borel_defBS}), reproduces the original function $f (z)$ when $f(z)$ is analytic as demonstrated below in the quadratic case.\footnote{
Note that the convergence of a perturpative series is not sufficient to reproduce the original function by Borel resummation.
For example, when $f(z)$ is an analytic function plus $e^{-1/z}$, the Borel resummation misses the latter part.
This kind of behaviour sometimes appears in supersymmetric systems \cite{Russo:2012kj,Aniceto:2014hoa,Honda:2016mvg,Dunne:2016jsr,Kozcaz:2016wvy,Dorigoni:2017smz,Dorigoni:2019kux}.
}
In this case, $f(z)$ is the same as a simple analytic continuation of the perturbative series summed inside its convergence radius, and correspondingly the Borel transformation, Eq.~\eqref{eq:Borel_defBT}, has an infinite radius of convergence allowing us to exchange the order of the integral and series expansion in Eq.~(\ref{eq:Borel_defBS}).
On the other hand, if $f (z)$ gives a divergent series but its Borel resummation is convergent, then $f_{\mathrm{S}} (z)$ is a function that has the same asymptotic behaviour as $f (z)$ {\it and} is convergent (in some angular domain in $z$-plane). 
In this sense, the Borel resummation endows the original formal series with an analytical meaning.
However, one or more singular points may appear along the contour of the integral ({\it i.e.} the real $t$ axis), and in such cases uncertainties arise as to how to avoid them. 
These ambiguities are typically related to non-perturbative aspects of the physical system.
However, we will see below that the correlation functions considered in the present paper have no such singularities and thus are free from uncertainties, allowing for unambiguous resummation.

Before applying Borel resummation to the divergent series of the quartic potential, let us demonstrate how it works for an exactly solvable case, the stochastic spectator in the quadratic potential. 
One starts with the original series~(\ref{eq:stochinf_expvquad}), 
\begin{equation}
    \expval{ \bar{ \phi }^2 } (\bar{ N }) 
    = \sum_{k = 0}^{\infty} \bar{a}_{2, k} \bar{N}^k 
    = \sum_{k = 1}^{\infty} \bar{a}_{2, k} \bar{N}^k 
    = \sum_{k = 0}^{\infty} \bar{a}_{2, k+1} \bar{N}^{k+1} \,\, , 
\end{equation}
where we used $\bar{a}_{2, 0} = 0$, 
then the Borel transformation of Eq.~(\ref{eq:stochinf_expvquad}) is obtained as 
\begin{equation}
  \expval{ \bar{ \phi }^2 }_{\mathrm{B}} (t) 
  = \sum_{k = 0}^{\infty} \frac{ a_{2, k+1} }{ \Gamma (k + 1) } t^k 
  = 
  \eval{ 
    \frac{1}{2 \pi^2} \frac{ I_1 (2 s) }{2 s} 
    }_{ s^2 = - 2 t / 3 }
  \,\, , 
  \label{eq:Borel_expvquadB}
\end{equation}
where $I_{\nu} (z)$ is the modified Bessel function of the first kind, and we used 
\begin{equation}
  I_{\nu} (z) 
  = \sum_{k = 0}^{\infty} 
  \frac{1}{k! \Gamma (\nu + k + 1)} \qty( \frac{z}{2} )^{\nu + 2k} \,\, . 
  \label{eq:Borel_Bessel1}
\end{equation}
Note that $t$ appearing in Eq.~(\ref{eq:Borel_expvquadB}) is just an auxiliary variable and has nothing to do with the time variable.
The absence of singularity in the Borel transformation~(\ref{eq:Borel_expvquadB}) in $t \in [0, \, \infty)$ implies that the system is free from non-perturbative effects and that the succeeding Laplace integral can be performed without any ambiguity.
One sees that the Borel transformation~(\ref{eq:Borel_expvquadB}) vanishes as $t \to \infty$, 
and this behaviour guarantees the relaxation of the correlators of the stochastic field that obeys the Langevin equation.
From Eq.~(\ref{eq:Borel_expvquadB}) we obtain the Borel resummation of $\expval{ \phi^2 }$ as
\begin{equation}
  \expval{ \bar{ \phi }^2 }_{\mathrm{S}} ( \bar{N} ) 
  = \int_{0}^{\infty} \dd t \, e^{-t / \bar{N}} \expval{ \bar{\phi}^2 }_{\mathrm{B}} (t) 
  = \frac{3}{8 \pi^2} \qty[
    1 - \exp \qty( - \frac{2}{3} \bar{N} )
  ] \,\, .
  \label{eq:Borel_resumquad}
\end{equation}
Here we used the identity~\cite{Gradshteyn:1702455}, 
\begin{equation}
  \int_{0}^{\infty} \dd z \, e^{- \alpha z^2} I_{\nu} (\beta z) = \frac{1}{2} \sqrt{ \frac{\pi}{\alpha} } \exp \qty( \frac{\beta^2}{8 \alpha} ) I_{\nu / 2} \qty( \frac{\beta^2}{8 \alpha} ) \,\, , 
  \quad 
  \Re \nu > -1 \,\, , 
  \quad 
  \Re \alpha > 0 \,\, . 
  \label{eq:Borel_Lap1}
\end{equation} 
As we see from this example, when the original function is an entire function (more generally analytic function), the Borel transformation is free from singularities everywhere and we can safely perform the Laplace integral to reproduce the original function exactly.

\subsection{Borel--Pad\'{e} resummation for a stochastic spectator in the quartic potential}
\label{sec:BorelPade}

\begin{figure}
  \begin{minipage}[b]{0.495\linewidth}
    \centering
    \subcaption{
        Borel--Pad\'{e} transf. for $V (\phi) = \lambda \phi^4 / 4$.
    }
    \includegraphics[width = 0.95\linewidth]{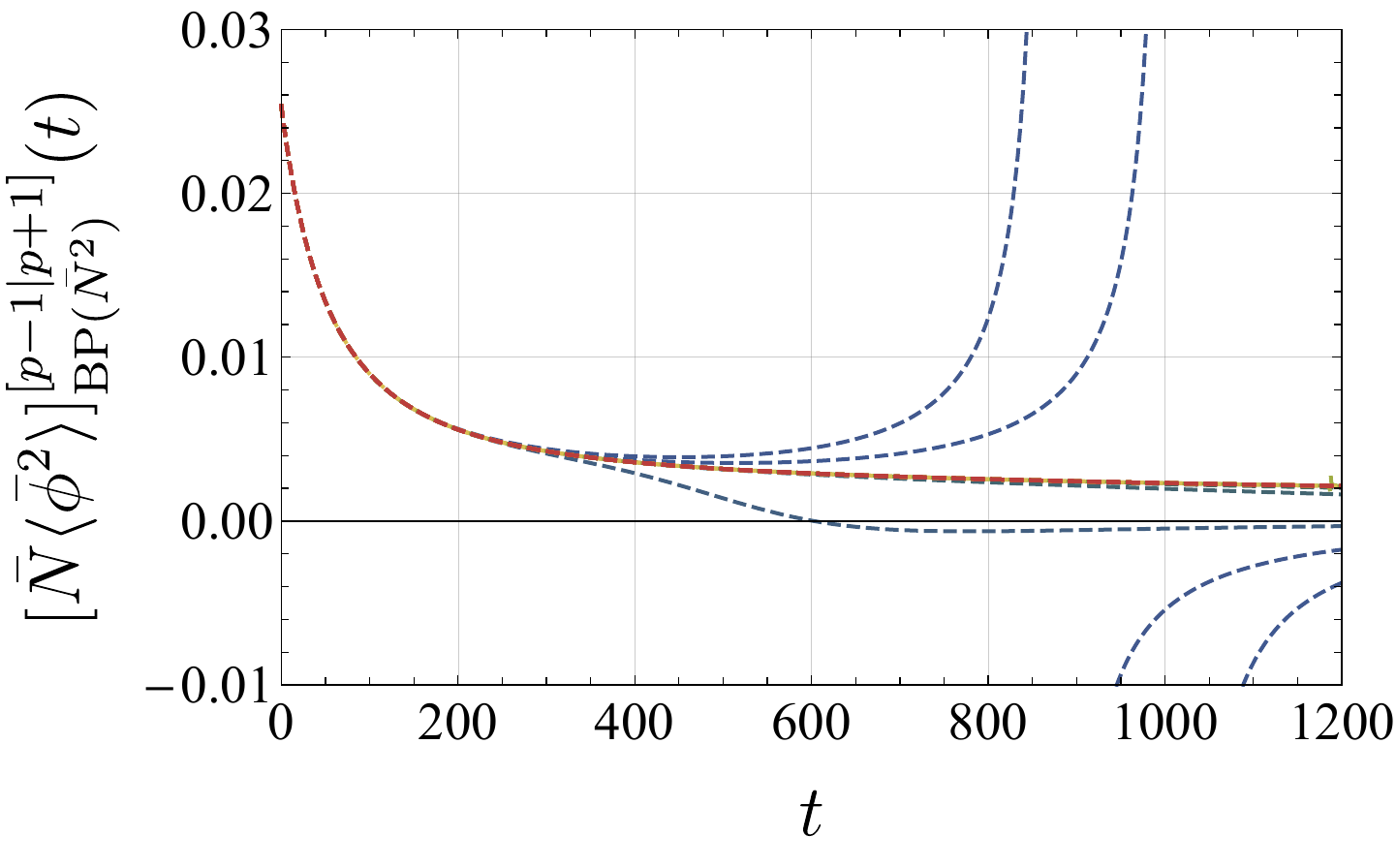}
  \end{minipage}
  \begin{minipage}[b]{0.495\linewidth}
    \centering
    \subcaption{
        Borel--Pad\'{e} transf. for $V (\phi) = \lambda \phi^4 / 4$.
    }
    \includegraphics[width = 0.95\linewidth]{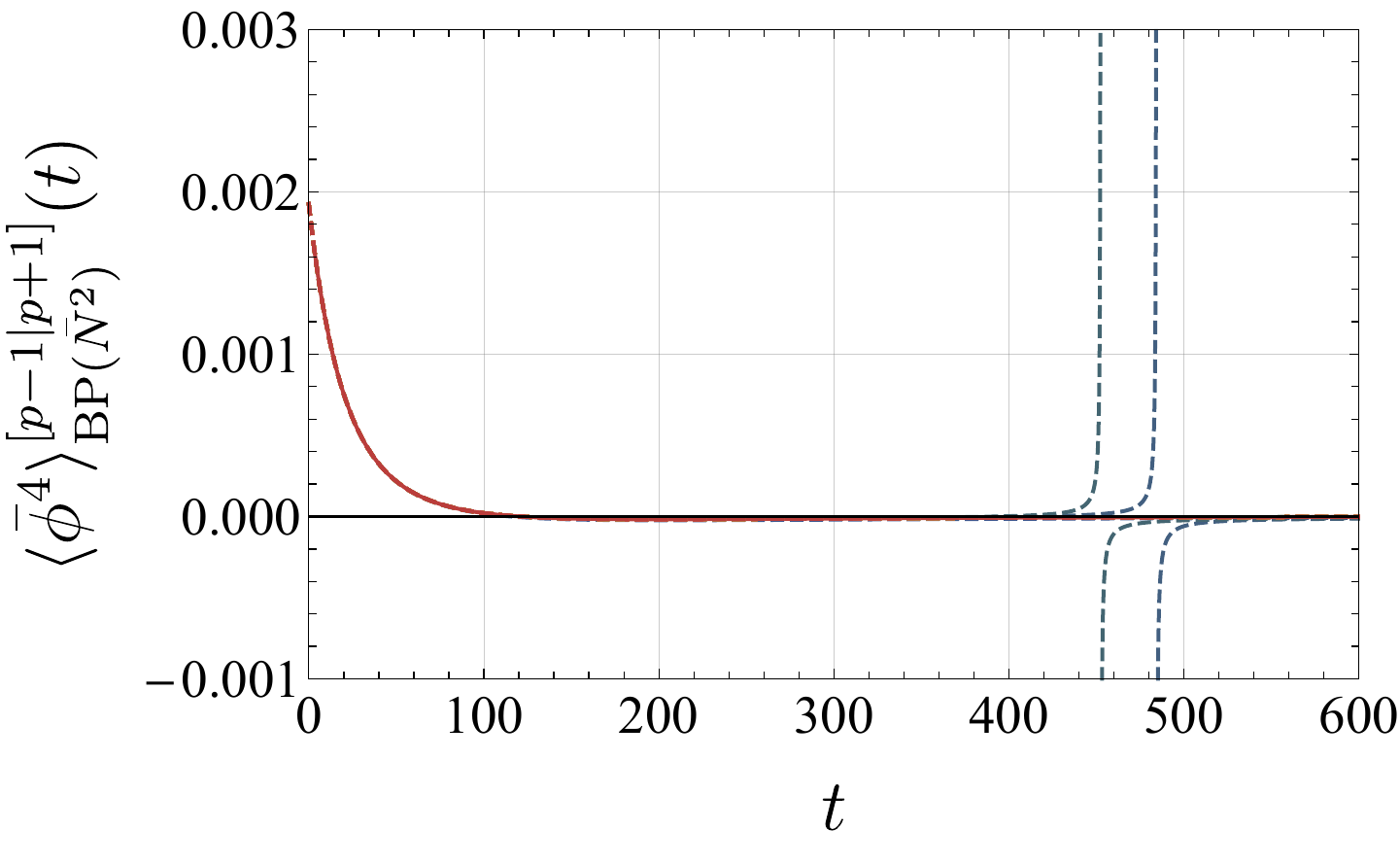}
  \end{minipage}
  \caption{
        Borel--Pad\'{e} transformation $\qty[ \bar{N} \expval{\bar{\phi}^2} ]_{\mathrm{BP} (\bar{N}^2)}^{[p - 1 | p + 1]} (t)$ (\textit{left}) and $\expval{\bar{\phi}^4}_{\mathrm{BP} (\bar{N}^2)}^{[p - 1 | p + 1]} (t)$ (\textit{right}) for $V = \lambda \phi^4 / 4$.
    The orders are $p = 2, \, 3, \, \dots, \, 30$, and $p$ increases from blue to red. 
    Some of the curves are degenerate.
  }
   \label{fig:BorelPade_trf}
\end{figure}

Let us apply Borel resummation to the spectator in the quartic potential. 
Since the expansion coefficients $\bar{a}_{2, k}$ for even $k$'s vanish, we may remove these coefficients,
\begin{equation}
    \expval{ \bar{ \phi }^2 } ( \bar{ N } ) 
    = \sum_{ k = 0 }^{\infty} \bar{a}_{2, k} \bar{N}^k 
    = \sum_{ \ell = 0 }^{\infty} \bar{a}_{2, 2 \ell + 1} \bar{N}^{2 \ell + 1} \,\, . 
    \label{eq:quarBorel_orig}
\end{equation}
In order to apply Eq.~(\ref{eq:Borel_series}) with $\bar{N}^2$ being the expansion parameter, we regard Eq.~(\ref{eq:quarBorel_orig}) as 
\begin{equation}
    \bar{N} \expval{ \bar{ \phi }^2 } ( \bar{ N } ) 
    = \sum_{ \ell = 0 }^{\infty} \bar{a}_{2, 2 \ell + 1} ( \bar{N}^2 )^{\ell + 1} \,\, . 
    \label{eq:quarBorel_sq}
\end{equation}
Borel transformation is applied to Eq.~(\ref{eq:quarBorel_sq}), 
\begin{equation}
    \qty[ \bar{N} \expval{ \bar{ \phi }^2 } ]_{\mathrm{B} (\bar{N}^2)} (t) 
    = \sum_{k = 0}^{\infty} \frac{ \bar{a}_{2, 2 k + 1} }{ k! } t^{k} \,\, . 
    \label{eq:quarBorel_sqBT}
\end{equation}
Note that the subscript $\mathrm{B} (\bar{N}^2)$ indicates that we perform Borel transformation with $\bar{N}^2$ being the expansion parameter.
The Laplace integral of Eq.~(\ref{eq:quarBorel_sqBT}) gives the Borel summation of $ \bar{N} \expval{ \bar{ \phi }^2 }$. Then, the Borel summed correlator reads 
\begin{equation}
    \expval{ \bar{ \phi }^2 }_{\mathrm{S} (\bar{N}^2)} (\bar{N}) 
    = \frac{1}{\bar{N}} \int_{0}^{\infty} \dd t \, e^{- t / \bar{N}^2} \qty[ \bar{N} \expval{ \bar{ \phi }^2 } ]_{\mathrm{B} (\bar{N}^2)} (t) \,\, . 
\end{equation}
The above procedure gives the Borel resummation for the formal series~(\ref{eq:stochinf_expvquar}) if all the coefficients $\bar{a}_{2, k}$ are available. 
However, in the present case, it is practically impossible to have all of them as we saw in Sec.~\ref{sec:section2}.
In this situation, one of the standard prescription is to approximate the Borel transformation with a Pad\'{e} approximant and then perform Laplace transformation.
This method is called \textit{Borel--Pad\'{e} resummation}.
First, the original series is truncated at a finite order, and from it the (truncated) Borel transformation is constructed. 
The Pad\'{e} approximant is used here giving the Borel--Pad\'{e} transformation of $\bar{N} \expval{ \bar{ \phi }^2 } (\bar{N})$,
\begin{equation}
  \qty[ \bar{N} \expval{ \bar{\phi}^2 } ]_{\mathrm{BP} (\bar{N}^2)}^{[p - 1 | p + 1]} (t) 
  = \qty[ 
    \sum_{k = 0}^{\infty} \frac{\bar{a}_{2, 2k + 1}}{k!} t^k 
  ]_{\mathrm{P}}^{[p - 1 | p + 1]} \,\, ,
\end{equation}
and we finally obtain the Borel--Pad\'{e} resummation,
\begin{equation}
  \expval{ \bar{ \phi }^2 }_{\mathrm{SP} (\bar{N}^2)}^{[p - 1 | p + 1]} (\bar{N}) 
  = \frac{1}{\bar{N}} \int_{0}^{\infty} \dd t \, e^{-t / \bar{N}^2} \qty[ \bar{N} \expval{ \bar{ \phi }^2 } ]_{\mathrm{BP} (\bar{N}^2)}^{[p - 1 | p + 1]} (t) \,\, . 
  \label{eq:bp_lap}
\end{equation}
In general, a Pad\'{e} approximant has one or more poles since it is a rational function by definition. 
Some of them are apparent ones that can (dis)appear depending on the choice of the order $p$, while others are manifestation of the singularities that the \textit{exact} Borel transformation has.
As we will see in Sec.~\ref{subsec:sing_Borel_plane}, the Borel--Pad\'{e} transformations have no poles on the positive real axis (except for some apparent ones, see for example the blue curves in Fig.~\ref{fig:BorelPade_trf}). 
Hence, whenever the Borel--Pad\'{e} transformations $\qty[ \bar{N} \expval{ \bar{\phi}^2 } ]_{\mathrm{BP} (\bar{N}^2)}^{[p-1|p+1]} (t)$ and $\expval{ \bar{\phi}^4 }_{\mathrm{BP} (\bar{N}^2)}^{[p-1|p+1]} (t)$ at some order $p$ have those apparent poles on the integration contour, we evaluate the Laplace integral~(\ref{eq:bp_lap}) taking the principal values at these poles.

\begin{figure}[!h]
    \begin{minipage}[b]{0.95\linewidth}
        \centering
        \subcaption{
            Borel--Pad\'{e} resummation for $V (\phi) = \lambda \phi^4 / 4$. 
        }
            \includegraphics[clip, width = 0.88\linewidth]{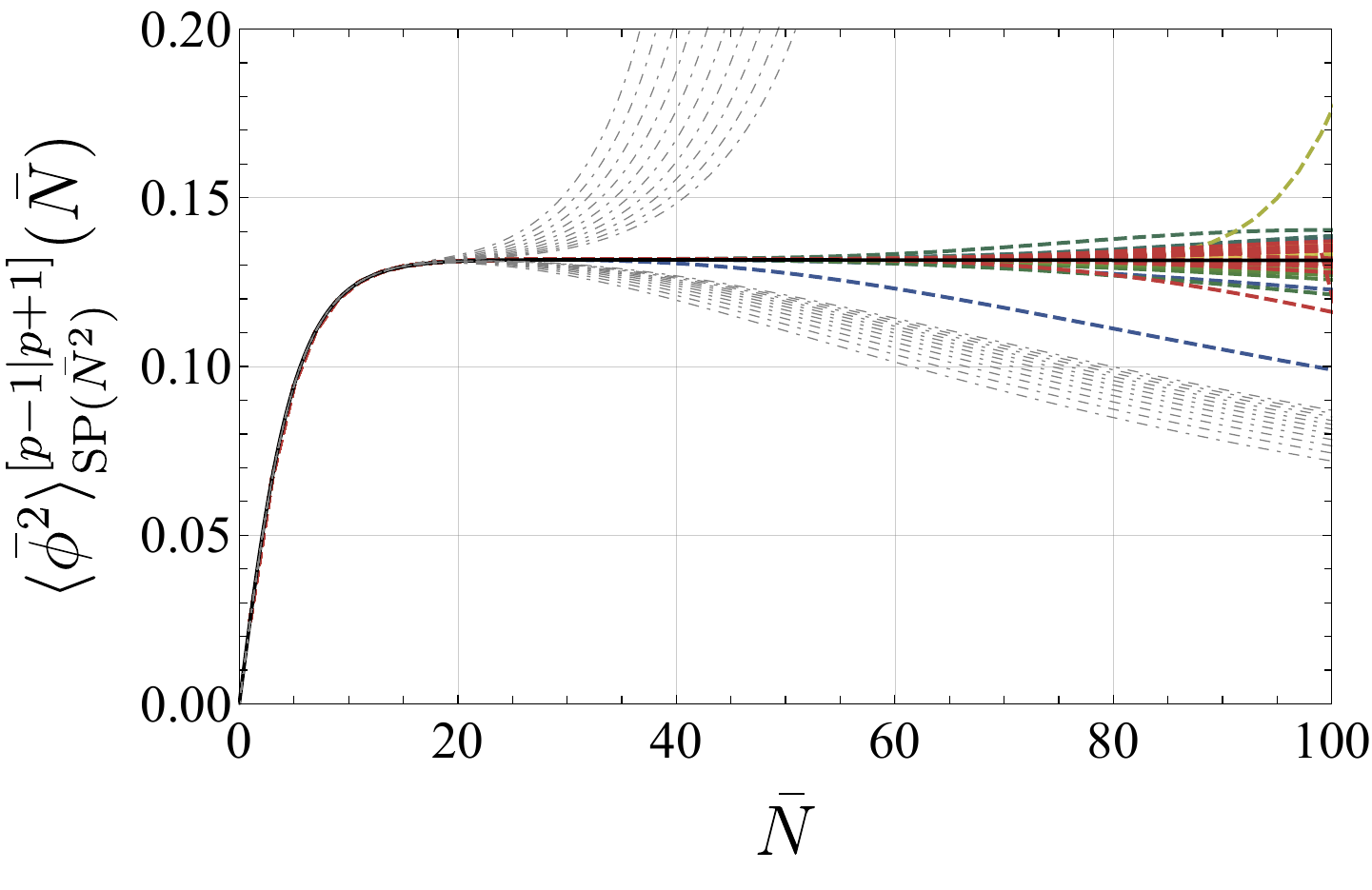}
    \end{minipage}
    \\[2.0ex]
    \begin{minipage}[b]{0.95\linewidth}
        \centering
        \subcaption{
            Borel--Pad\'{e} resummation for $V (\phi) = \lambda \phi^4 / 4$.
        }
            \includegraphics[clip, width = 0.88\linewidth]{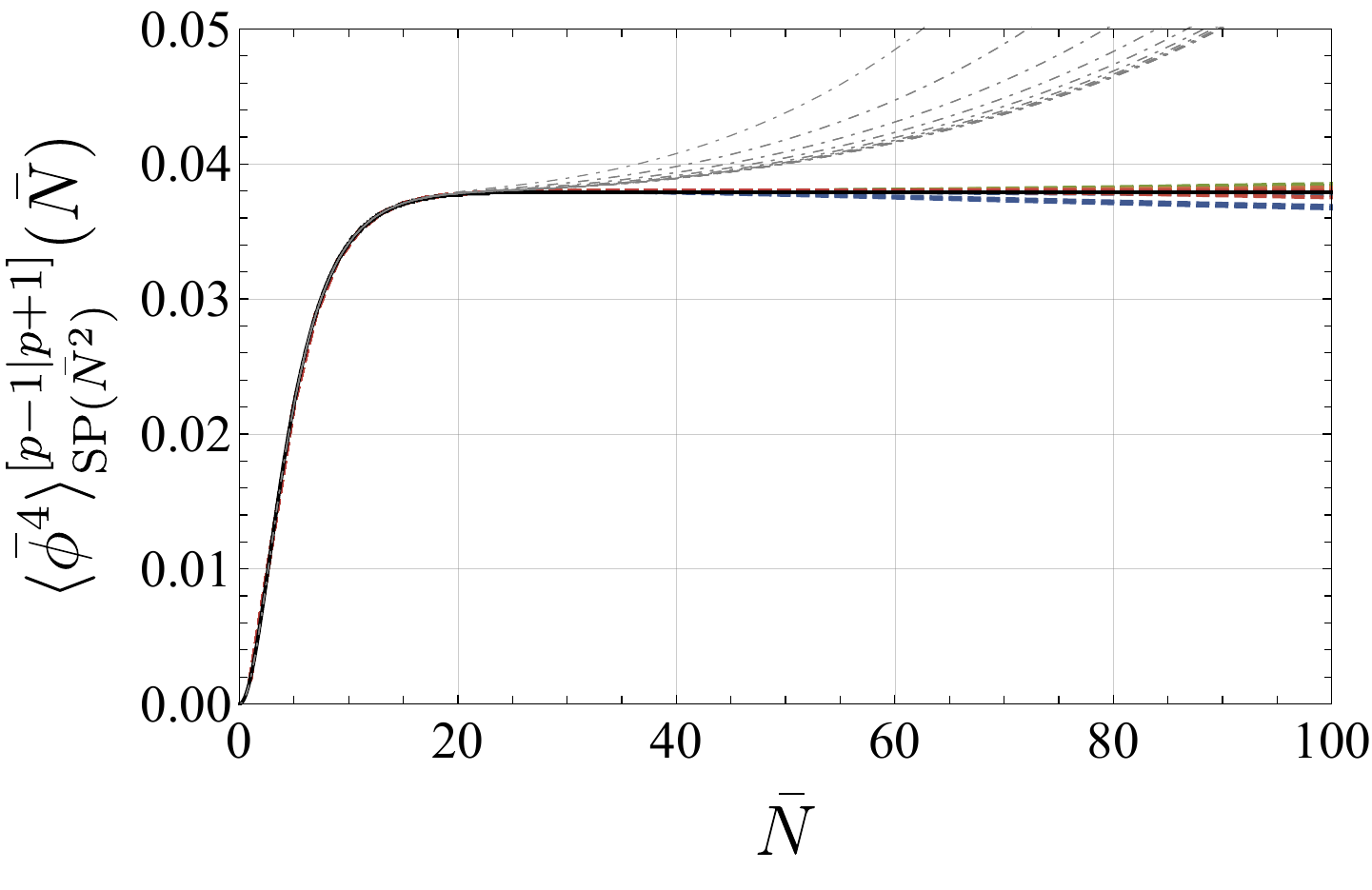}
    \end{minipage}
    \label{fig:label-ABC}
    \caption{
        The Borel--Pad\'{e} resummation $\expval{\bar{\phi}^2}_{\mathrm{SP} (\bar{N}^2)}^{[p - 1 | p + 1]} (\bar{N})$ (\textit{left}) and $\expval{\bar{\phi}^4}_{\mathrm{SP} (\bar{N}^2)}^{[p - 1 | p + 1]} (\bar{N})$ (\textit{right}) for $V = \lambda \phi^4 / 4$. 
        The orders are $p = 16, \, \dots, \, 50$, and $p$ increases from blue to red, while the black solid lines correspond to the numerically calculated behaviour. 
        In the right panel, most of all the curves are degenerate.
        The direct Pad\'{e} approximants for $p = 30, \, \dots, \, 50$ are also plotted with grey dotted-dashed curves for comparison. 
    }
    \label{fig:BorelPade}
\end{figure}

Figure~\ref{fig:BorelPade_trf} shows the Borel--Pad\'{e} transformation at different orders $p$ in the Laplace space (\textit{i.e.} as a function of $t$).
The left and right panels are $\qty[ \bar{N} \expval{ \bar{\phi}^2 } ]_{\mathrm{BP} (\bar{N}^2)}^{[p - 1 | p + 1]} (t) $ and $\expval{ \bar{\phi}^4 }_{\mathrm{BP} (\bar{N}^2)}^{[p - 1 | p + 1]} (t) $, respectively. 
Figure~\ref{fig:BorelPade} shows the result of the Borel--Pad\'{e} resummation, with the left and right panel being $\expval{ \bar{ \phi }^2 }_{\mathrm{SP} (\bar{N}^2)}^{[p - 1 | p + 1]}(\bar{N})$ and $\expval{ \bar{ \phi }^4 }_{\mathrm{SP} (\bar{N}^2)}^{[p - 1 | p + 1]}(\bar{N})$, respectively.
For comparison, we also show the result of the direct Pad\'{e} in grey.
We see that both the transient and stationary behaviour are nicely reproduced, and that the Borel--Pad\'{e} improves the approximation compared to the direct Pad\'{e} in Fig.~\ref{fig:DirectPade}.
This is the main result of this paper.

Note that Eq.~(\ref{eq:quarBorel_sq}) regards $\bar{N}^2$ as the expansion parameter rather than $\bar{N}$.
When the initial condition for the spectator field is taken arbitrary, one cannot necessarily regard the former to be the expansion parameter since the coefficients $\bar{a}_{n, k}$ may have non-zero entries for both odd and even orders of $k$.
We show in Appendix~\ref{app:appendix1} that Borel--Pad\'{e} transformation works even in such cases. 

\subsection{Singularity structure in the Borel plane}
\label{subsec:sing_Borel_plane}

In this subsection we finally study the singularity structure in the Borel plane for $V (\phi) = \lambda \phi^4 / 4$.
The singularity structure is important for the following reasons.
First, the location of the singularities affects whether the Borel resummation is well-defined: the integral contour of the Laplace transformation may hit the singularities and hence we should check if it happens.
Second, it is known that singularities of the Borel transformation are typically related to non-perturbative effects and Stokes phenomena.
While we do not have an exact expression for the Borel transformation in the current problem, it is natural to expect that the Borel--Pad\'{e} transformation reflects the original singularity structure to some extent.
Thus in the following we estimate it through the Borel--Pad\'{e} transformation.

\begin{figure}
\centering
\includegraphics[clip, width = 0.95\linewidth]{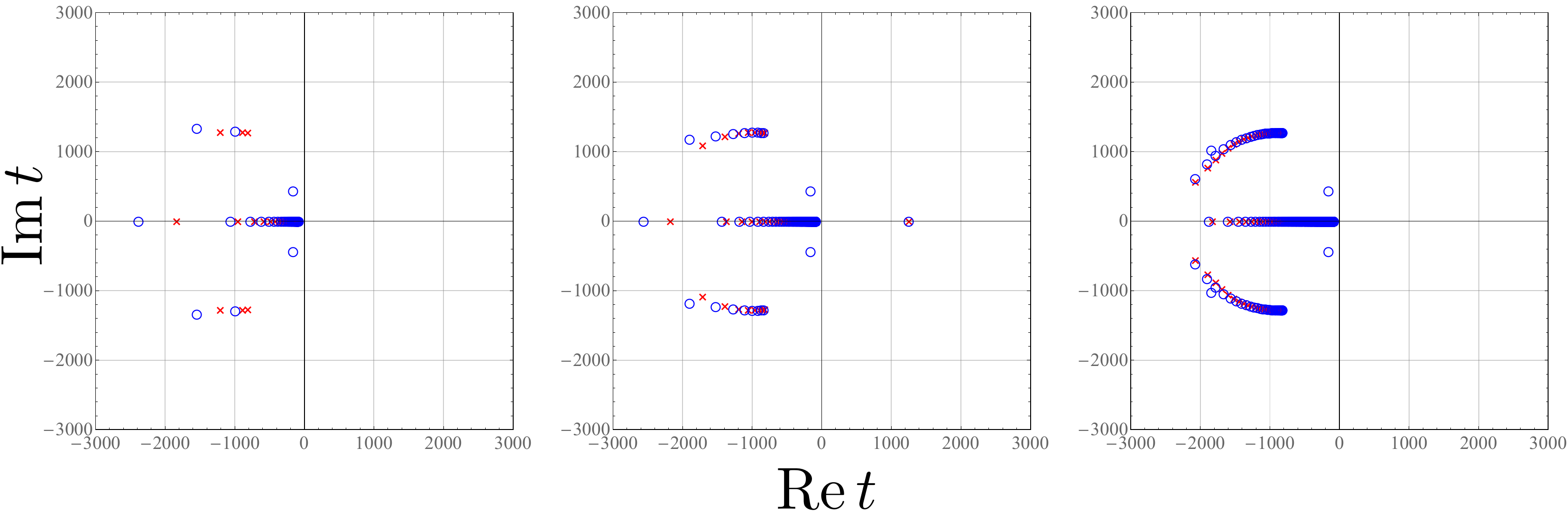}
\\[2.0ex]
\includegraphics[clip, width = 0.95\linewidth]{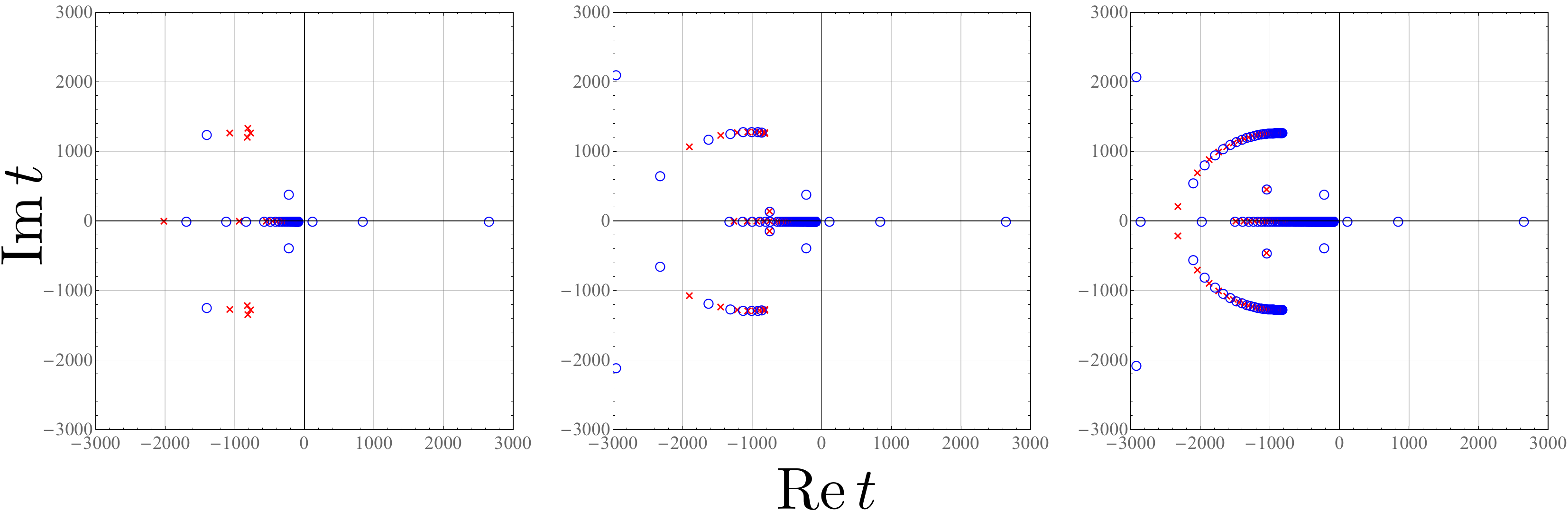}
\caption{
    Distribution of the poles and zeros in the Borel--Pad\'{e} transformation, $\qty[ \bar{N} \expval{\bar{\phi}^2} ]_{\mathrm{BP} (\bar{N}^2)}^{[p - 1 | p + 1]} (t)$ (\textit{top}) and $\expval{\bar{\phi}^4}_{\mathrm{BP} (\bar{N}^2)}^{[p - 1 | p + 1]} (t)$ (\textit{bottom}), for $V = \lambda \phi^4 / 4$. 
    Plots are made for $p = 50$ (\textit{left}), $p = 150$ (\textit{middle}), and $p = 500$ (\textit{right}). 
    The appearance of alternating poles (red crosses) and zeros (blue circles) along the curves imply the existence of branch cuts in the \textit{exact} Borel transformation.
}
\label{fig:sing_singularity}
\end{figure}

Figure~\ref{fig:sing_singularity} shows how the poles (red crosses) and zeros (blue circles) in the Borel--Pad\'{e} transformation of $\bar{N} \expval{ \bar{\phi}^2 } (\bar{N})$ and $\expval{ \bar{\phi}^4 } (\bar{N})$ change as the order $p$ increases. 
We observe several clusters in which poles and zeros appear alternately:
one is located along the negative real axis, and the others form curves in the left half of the $t$-plane.
In all the three curves the poles and zeros appear alternately, and the three curves are relatively stable against the change in the order of the Pad\'{e} approximant.
These facts suggest that these poles and zeros inherit the branch cuts that the \textit{exact} Borel transformation has~\cite{yamada2014numerical}. 
One of the cuts lies along the negative real axis, and it starts from $t = t_0 \simeq - 80$.
The starting point $t = t_0$ determines the convergence radius of the series expansion of the exact Borel transformation. 
The others extend to the real negative axis as the order $p$ increases.
The existence of the cuts signals that there are Stokes phenomena when we extend $\bar{N}^2$ to complex region.
However, practically this is not of much importance: what is important here is that we do not have cuts extended to the real and positive axis nor isolated poles on it, and thus the succeeding Laplace integral has no ambiguity arising from the way to circumvent the branch cuts or poles on the integration contour. 
This also suggests the absence of non-perturbative effects in the present system.

Other structures include isolated zeros that appear in all the panels in Fig.~\ref{fig:sing_singularity}.
However, zeros do not mean any singularities and thus they do not have much importance in the current analysis.
Also, in identifying the location of the poles and zeros, care must be taken with numerical precision since the coefficients of the perturbative series are calculated with a finite numerical precision. 
Insufficient precision can lead to the emergence of {\it ghost pairs} in a characteristic way~\cite{yamada2014numerical}, and we explain this phenomenon in Appendix~\ref{app:appendix2}. 

\section{Discussion and conclusions}
\label{sec:section4}

In this paper, we discussed the application of the Pad\'{e} approximation and Borel--Pad\'{e} resummation in the context of the stochastic formalism of inflation, focusing on the dynamics of a spectator field.
In the stochastic formalism, coarse-grained fields follow the Langevin equation and their distributions correspondingly follow the Fokker-Planck equation.
In de Sitter spacetime, it is intuitively easy to understand that this is a system leading to an equilibrium state. 
In fact, the equilibrium distribution (\ref{eq:stochinf_stat}) and (\ref{eq:stochinf_statexpv}) can be easily calculated for an arbitrary potential, and the relaxation process is not difficult to obtain numerically (see Fig.~\ref{fig:eqb_pdf}).
However, it is hard to grasp an analytical understanding of the out-of-equilibrium transition.
One way is to perturbatively expand the correlation functions in terms of the $e$-folding number $N$.
But the series are dangerously diverging for a wide class of potentials and can only be interpreted as formal series.
They are therefore reputed to be trustworthy at early times only.

To investigate the properties of the correlation functions in stochastic inflation and the usefulness of the Pad\'{e} approximation or the Borel--Pad\'{e} resummation, we focused on the dynamics of spectator field in the paradigmatic $\lambda \phi^4$ setup in de Sitter spacetime.
First, we confirmed that the expansion coefficients of the correlation functions increase more or less factorially with an alternating sign, suggesting that the radius of convergence of the series is zero (see the right panel of Fig.~\ref{fig:coefficients_p2} and the bottom panels of Fig.~\ref{fig:truncation}). 
This is in contrast to the case of a quadratic drift, in which the system is analytically solvable and the coefficients monotonically decrease (see the left panel of Fig.~\ref{fig:coefficients_p2}).
The Pad\'{e} approximants may be useful in taming such diverging behaviour, and we first explored this possibility.
They approximate a function by a rational function in such a way that the coefficients are determined so that the power series of the latter successively matches that of the former, and in many cases they reproduce the original function better than the naively truncated series.
While the naively truncated series gets worse as the truncation order increases, the Pad\'{e} approximants indeed reproduce the original functions better as the order increases, up to a certain $e$-folding, and not only around the transient regime but also until the equilibrium is reached (see Fig.~\ref{fig:DirectPade}).

While the Pad\'{e} approximants are useful, we saw that they perform relatively worse in the quartic case presumably because this case has a divergent perturbative series and the Pad\'{e} approximant is likely worse at approximating such functions than analytic functions as appeared in the quadratic case.
This motivates us to apply the Borel--Pad\'{e} resummation, where the Borel transformation is approximated by the Pad\'{e} approximant.
There are two other reasons to use this method.
One is that the present system is an equilibrating one in which the correlation functions are expected to reach constant values, and in such a case, the Borel transformation converges to zero in the Borel plane as its argument $t$ goes to infinity.
Due to this property, we expect good accuracy of the Pad\'{e} approximant in Borel plane and the resulting Borel--Pad\'{e} resummation.
In fact, we confirmed that Borel--Pad\'{e} resummation reproduces very well the behavior of the original correlation functions from the transient regime to the equilibrium regime (see Fig.~\ref{fig:BorelPade}).
Another reason is the general expectation that the singularity structure in Borel plane tells us about non-perturbative properties of the original system.
Although the singularity structure of the Borel transformation is not strictly known until all the expansion coefficients are available, its Pad\'{e} approximation often inherits the original singularity structure.
We found several clusters of the poles and zeros in the Borel--Pad\'{e} transformations: one is along the real negative axis, and it stems from the convergence boundary.
It has poles and zeros appearing in an alternate way, hence implying the existence of a branch cut. 
The others are located on curves, and again have poles and zeros appearing alternately, thus signalling other branch cuts. 
However, as these singularities do not appear on the positive real axis, we do not expect any non-perturbative effects present in the current setup. 
Therefore, the Laplace integral, the final step of the Borel(--Pad\'{e}) resummation, can be performed without ambiguity.

We conclude by mentioning several possible applications of the analysis presented in this paper.
First, it would be straightforward to extend the currents results to higher order $n$-point correlation functions within the same setup.
A natural direction would then be to investigate to which extent one may reconstruct the full PDF in the relaxation process, from the time-dependent correlation functions.
Second, in our analysis, the spectator field was assumed to start from the global minimum of $\mathbb{Z}_2$-symmetric potentials, which greatly simplified the recurrence structure. 
It would be interesting to study more general initial conditions, in which case those simplifications could not occur. 
Third, another possibility would be to study more nontrivial potentials such as the double well potential leading to phase transitions in the early universe, and to investigate the relation between the singularity structure and non-perturbative effects.
Indeed, the one-loop correction to the instanton contribution to the correlation function can be seen in the stochastic framework.

Last but not least, the case of the stochastic field being the inflaton field and therefore leading the expansion of the universe, would be of great importance.
For this purpose one should take the field dependence of the Hubble parameter into account, as given by the Friedmann equation in the slow-roll approximation, $3H^2 M_\mathrm{Pl}^2 \simeq V(\phi)$.
One can already anticipate several other complications, such as the fact that the discretisation of the time scheme of the Langevin equations could affect the final result.
Then, since fluctuations of the inflaton can be converted to those of the curvature perturbation through the stochastic $\delta N$ formalism, 
one can treat the latter within the stochastic framework in a non-perturbative way.
One of the interesting consequences of this approach is that the PDF of the curvature perturbation typically develops an exponential tail, which can be expected to lead to a more efficient PBH formation scenario than the same setup investigated with the conventional linear perturbation theory. 
Understanding the exponential tail from the viewpoint of singularities in the Borel space is definitely a thrilling future direction.
We leave such considerations for future work.

\acknowledgments

The authors would like to thank Vincent Vennin for insightful comments on this manuscript.
M.~H. is supported by MEXT Q-LEAP, JST PRESTO Grant Number JPMJPR2117, JSPS Grant-in-Aid for Transformative Research Areas (A) JP21H05190 and JSPS KAKENHI Grant Number 22H01222.
L.~P. would like to acknowledge support from the “Ram\'{o}n y Cajal” grant RYC2021-033786-I, his work is partially supported by the Spanish Research Agency (Agencia Estatal de Investigaci\'{o}n) through the Grant IFT Centro de Excelencia Severo Ochoa No CEX2020-001007-S, funded by MCIN/AEI/10.13039/501100011033. 
K.~T. acknowledges the support from JSPS KAKENHI Grant No.~21J20818.

\appendix

\section{Different choices for Pad\'{e} approximant and Borel--Pad\'{e} summation}
\label{app:appendix1}

\paragraph{Direct Pad\'{e}}

\begin{figure}
  \begin{minipage}[b]{0.495\linewidth}
    \centering
     \subcaption{
        $V (\phi) = \lambda \phi^4 / 4$.
    }
    \includegraphics[width = 0.95\linewidth]{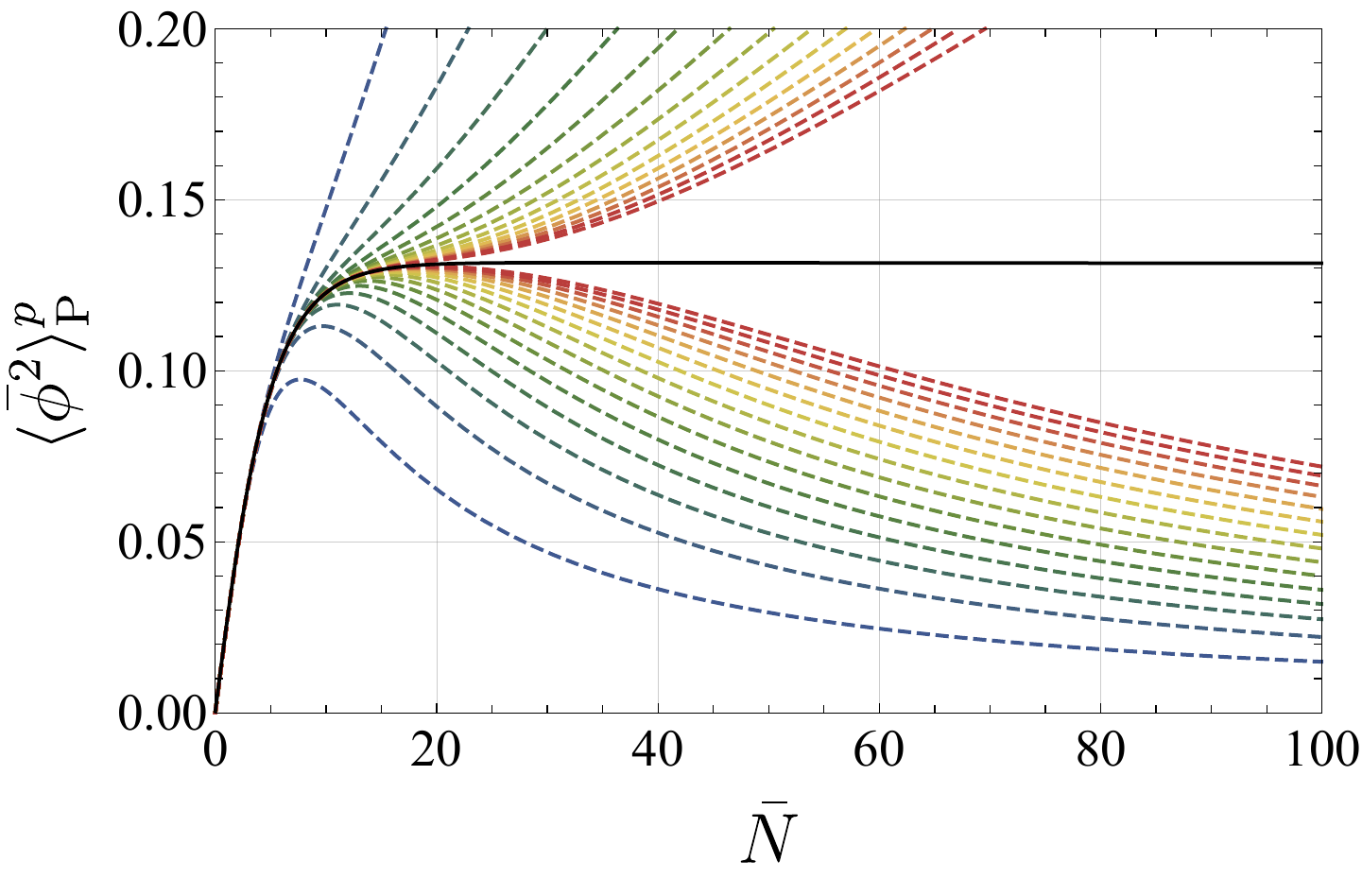}
  \end{minipage}
  \begin{minipage}[b]{0.495\linewidth}
    \centering
     \subcaption{
        $V (\phi) = \lambda \phi^4 / 4$.
    }
    \includegraphics[width = 0.95\linewidth]{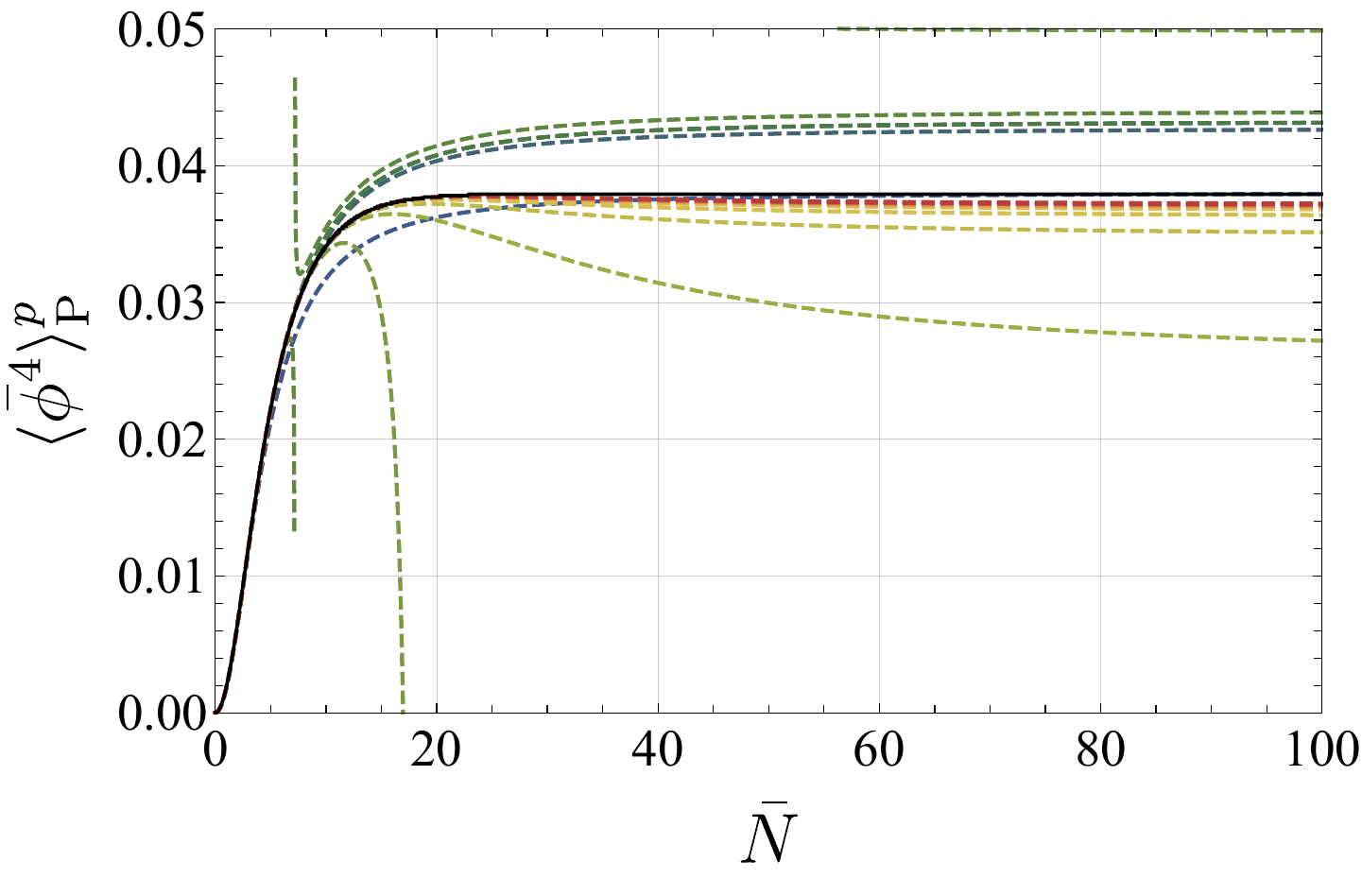}
  \end{minipage}
  \caption{
    Direct diagonal Pad\'{e} approximation  $\expval{\bar{\phi}^2}_{\textrm{P}}^{[p | p]} (\bar{N})$ (\textit{left}) and $\expval{\bar{\phi}^4}_{\textrm{P}}^{[p | p]} (\bar{N})$ (\textit{right}) for $V = \lambda \phi^4 / 4$. 
    The orders are $p = 2, \, \cdots, \, 30$ and $p$ increases from blue to red.
  }
  \label{fig:DirectPadeDiagonal}
\end{figure}

In the main text we used the non-diagonal Pad\'{e} $[p - 1 | p + 1]$ for the direct Pad\'{e} approximation.
However, since one knows that the correlators approach constant values for $N \to \infty$, one may wonder if diagonal Pad\'{e} performs better.
Actually this is true for $\expval{\phi^4}, \expval{\phi^8}, \cdots$ while not for $\expval{\phi^2}, \expval{\phi^6}, \cdots$.
In Fig.~\ref{fig:DirectPadeDiagonal} we show the results for the direct diagonal Pad\'{e}.
The left panel is for $\expval{\phi^2}_{\textrm{P}}^{[p | p]}$ while the right panel is for $\expval{\phi^4}_{\textrm{P}}^{[p | p]}$.
The former does not show significant improvement compared to the left panel of Fig.~\ref{fig:DirectPade}, while the latter improves drastically from the right panel of Fig.~\ref{fig:DirectPade}. 

The reason for the behavior of the correlators $\expval{\phi^2}, \expval{\phi^6}, \cdots$ can be explained in the following way.
Consider $\tanh x$, which approaches a constant for $x \to \infty$ and has only odd powers of the argument when expanded around zero:
\begin{align}
\tanh x
&=
x - \frac{1}{3} x^3 + \frac{2}{15} x^5 - \frac{17}{315} x^7 + \cdots.
\end{align}
Applying diagonal Pad\'{e} approximation, one obtains
\begin{align}
    [\tanh x]_{\mathrm{P}}^{[1|1]}
    &=
    x \,\, , 
    \quad 
    [\tanh x]_{\mathrm{P}}^{[2|2]}
    =
    \frac{x}{\displaystyle 1 + \frac{x^2}{3}} \,\, , 
    \quad 
    [\tanh x]_{\mathrm{P}}^{[3|3]}
    =
    \frac{\displaystyle x + \frac{x^3}{15}}{\displaystyle 1 + \frac{2 x^2}{5}} \,\, , 
    \quad 
    \cdots \,\, . 
\end{align}
As clear from these expressions, diagonal Pad\'{e} does not necessary mean that the highest order terms for the numerator and/or denominator are nonzero.
According to table \ref{tbl:phi4}, this class of correlators has only terms with odd powers of $N$ and thus cannot improved by the diagonal choice for the Pad\'{e} approximants.
\paragraph{Borel--Pad\'{e} resummation}
In Sec.~\ref{sec:section3}, we applied Borel transformation with $\bar{N}^2$ being the fundamental expansion parameter, see Eq.~(\ref{eq:quarBorel_sq}).
However, we may not always do this since the coefficients $a_{n,k}$ may have full entries depending on the initial condition for the spectator field.
In this appendix, therefore, we show how the results change for Borel--Pad\'{e} resummation with $\bar{N}$ being the expansion parameter.

\begin{figure}
    \centering
    \includegraphics[width = 0.95\linewidth]{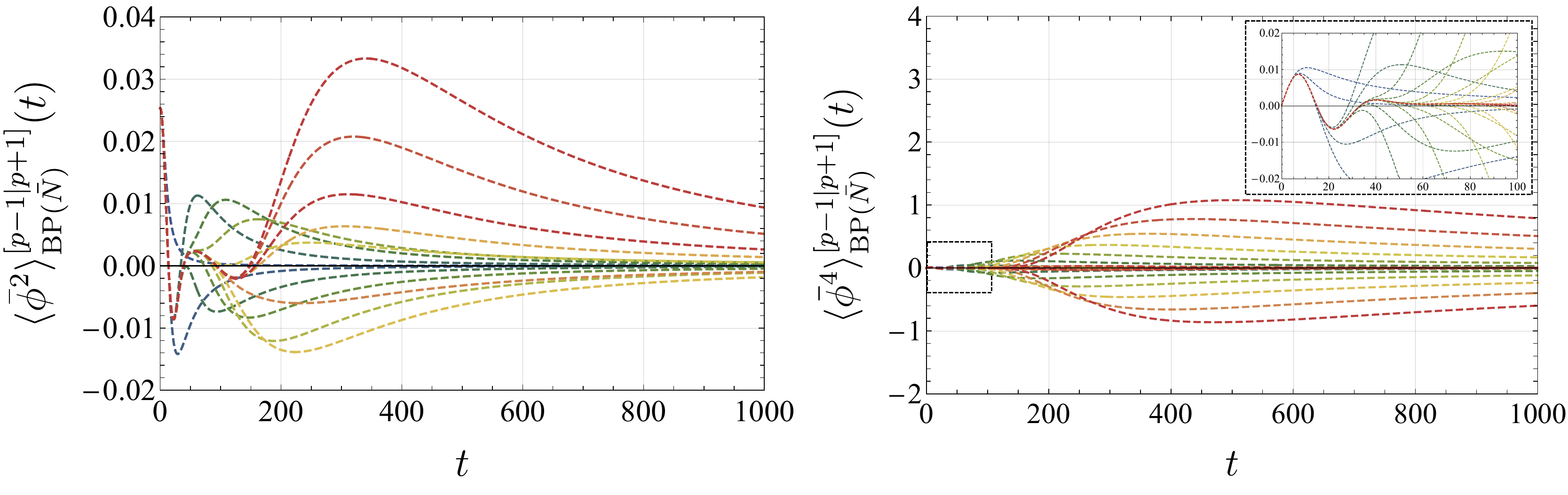}
    \caption{
    Borel--Pad\'{e} transformation $\expval{\bar{\phi}^2}_{\mathrm{BP} (\bar{N})}^{[p - 1 | p + 1]} (t)$ (\textit{left}) and $\expval{\bar{\phi}^4}_{\mathrm{BP} (\bar{N})}^{[p - 1 | p + 1]} (t)$ (\textit{right}) for $V = \lambda \phi^4 / 4$ with $\bar{N}$ being the expansion parameter.
    The orders are $p = 1, \, \dots, \, 30$ and $p$ increases from blue to red.
    }
    \label{fig:Quartic_NaiveBorelPadeTransf_N}
\end{figure}

\begin{figure}
  \begin{minipage}[b]{0.495\linewidth}
    \centering
     \subcaption{
        $V (\phi) = \lambda \phi^4 / 4$.
    }
    \includegraphics[width = 0.95\linewidth]{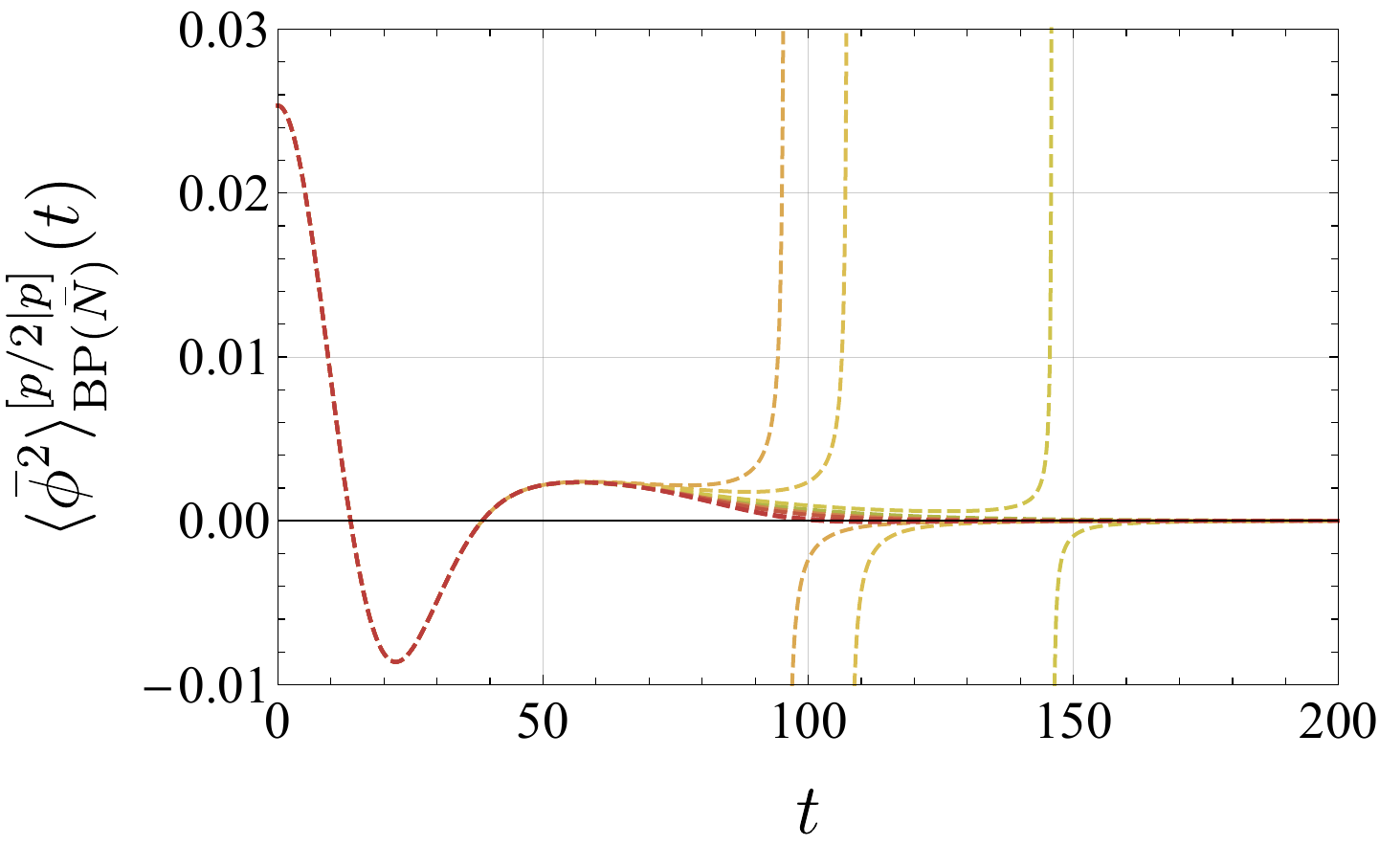}
  \end{minipage}
  \begin{minipage}[b]{0.495\linewidth}
    \centering
     \subcaption{
        $V (\phi) = \lambda \phi^4 / 4$.
    }
    \includegraphics[width = 0.95\linewidth]{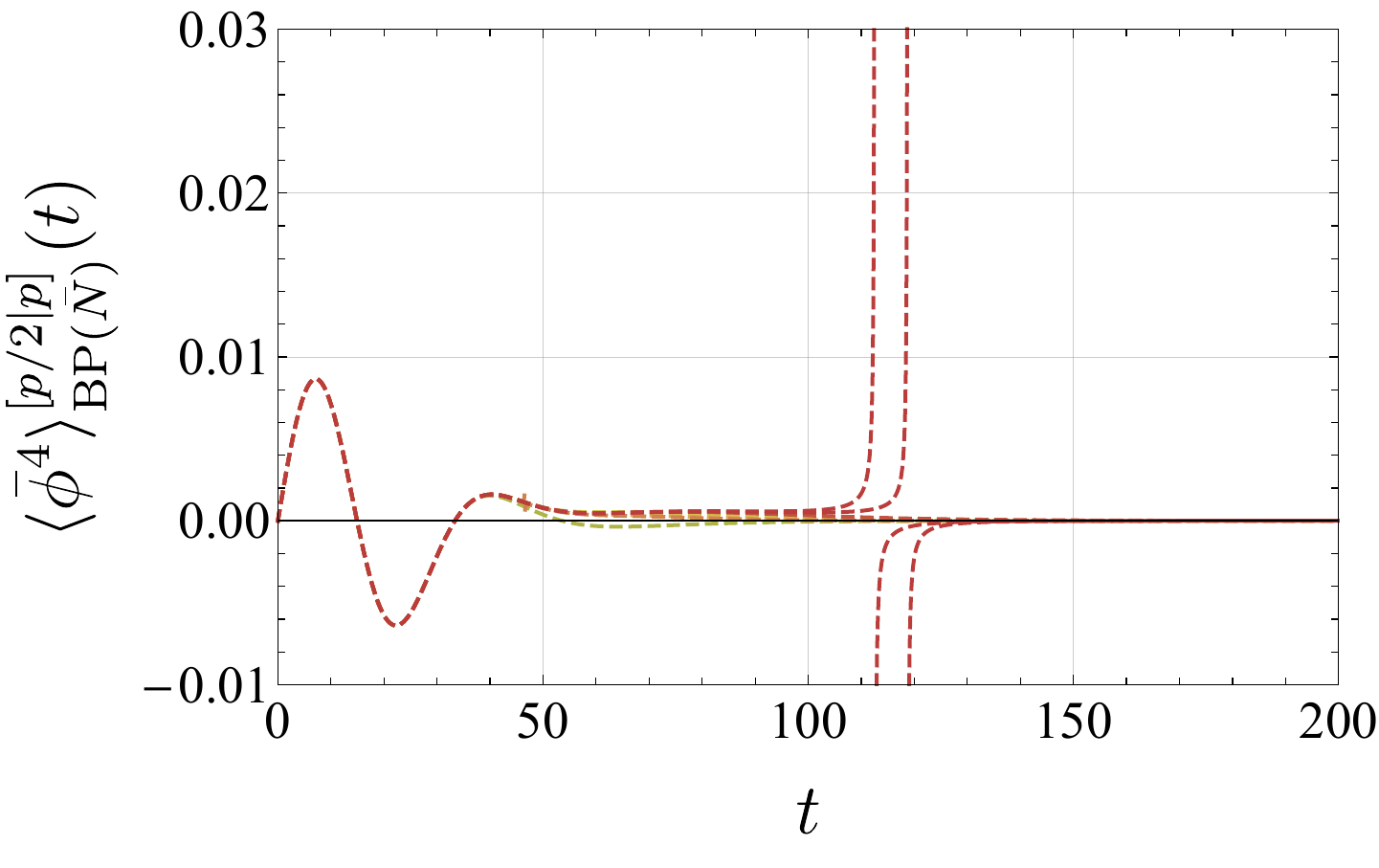}
  \end{minipage}
  \caption{
     Borel--Pad\'{e} transformation $\expval{\bar{\phi}^2}_{\mathrm{BP} (\bar{N})}^{[p/2 | p]} (t)$ (\textit{left}) and $\expval{\bar{\phi}^4}_{\mathrm{BP} (\bar{N})}^{[p/2 | p]} (t)$ (\textit{right}) for $V = \lambda \phi^4 / 4$ with $\bar{N}$ being the expansion parameter.
     The orders are $p = 16, \, 18, \, \dots, \, 30$ and $p$ increases from blue to red.
  }
  \label{fig:Quartic_BorelPadeTransf_N}
\end{figure}

We first show in Fig.~\ref{fig:Quartic_NaiveBorelPadeTransf_N} the Borel--Pad\'{e} transformation with $\bar{N}$ being the expansion parameter, using the same orders for the approximants as Sec.~\ref{sec:BorelPade}.
This corresponds to using the full coefficients $\bar{a}_{2, k + 1}$, not $\bar{a}_{2, 2 k + 1}$, and $[m | n] = [p - 1 | p + 1]$,
\begin{align}
  \expval{ \bar{\phi}^2 }_{\mathrm{BP} (\bar{N})}^{[p - 1 | p + 1]} (t) 
  &= \qty[ \sum_{k = 0}^{\infty} \frac{\bar{a}_{2, k + 1}}{k!} t^k 
  ]_{\mathrm{P}}^{[p - 1 | p + 1]} \,\, ,
  \\
  \expval{ \bar{\phi}^4 }_{\mathrm{BP} (\bar{N})}^{[p - 1 | p + 1]} (t) 
  &= \qty[ \sum_{k = 0}^{\infty} \frac{\bar{a}_{4, k + 1}}{k!} t^k 
  ]_{\mathrm{P}}^{[p - 1 | p + 1]} \,\, .
\end{align}
The Borel--Pad\'{e} transformation develops high peaks at large $t$ values, though the exact Borel transformation is expected to damp with oscillations (see also Fig.~\ref{fig:Quartic_BorelPadeTransf_N}).
These high peaks tend to spoil the asymptotic values of the Borel--Pad\'{e} summation when we go back to the $\bar{N}$-space via Laplace integral.

\begin{figure}
  \begin{minipage}[b]{0.495\linewidth}
    \centering
     \subcaption{
        $V (\phi) = \lambda \phi^4 / 4$.
    }
    \includegraphics[width = 0.95\linewidth]{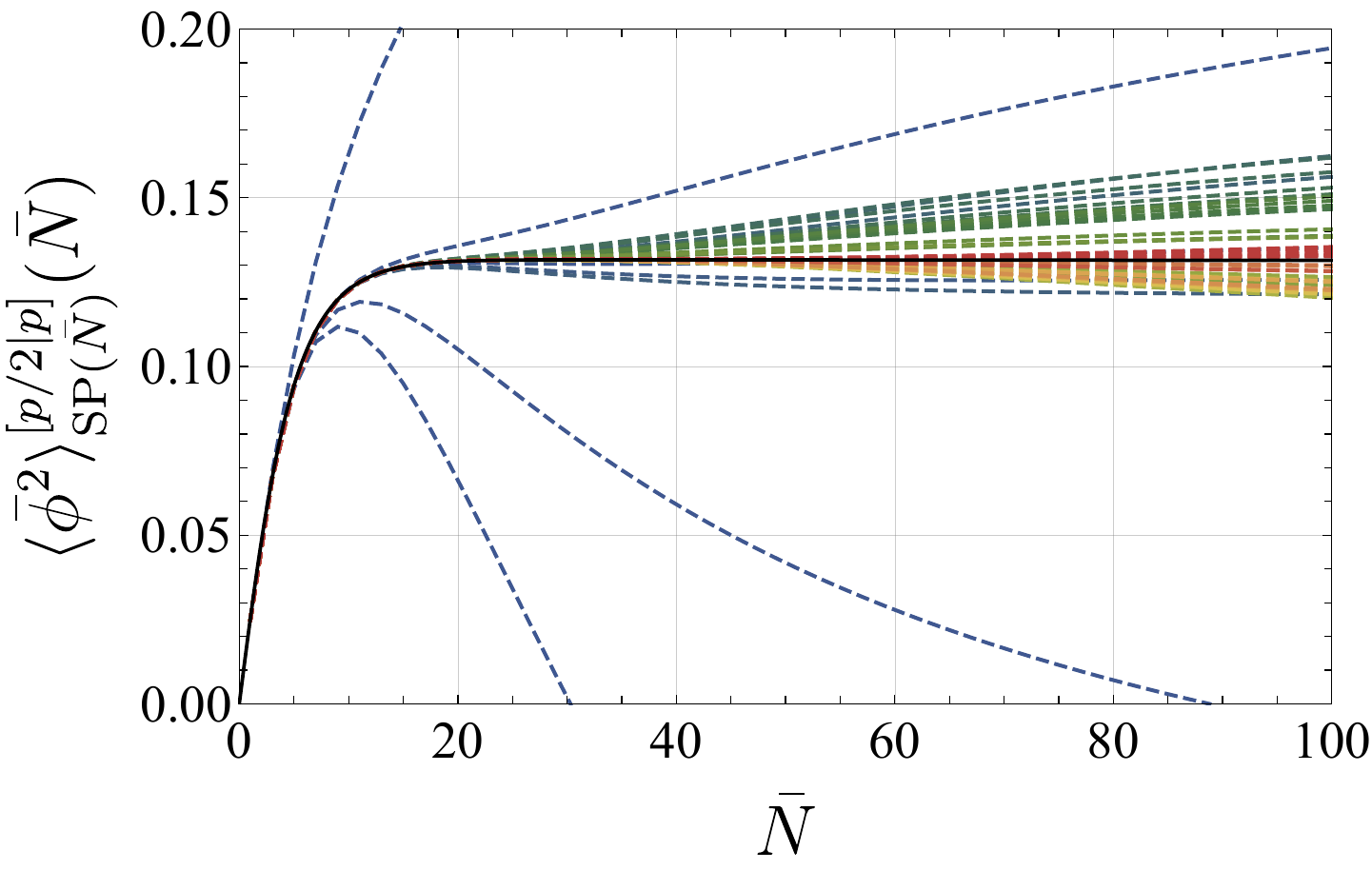}
  \end{minipage}
  \begin{minipage}[b]{0.495\linewidth}
    \centering
     \subcaption{
        $V (\phi) = \lambda \phi^4 / 4$.
    }
    \includegraphics[width = 0.95\linewidth]{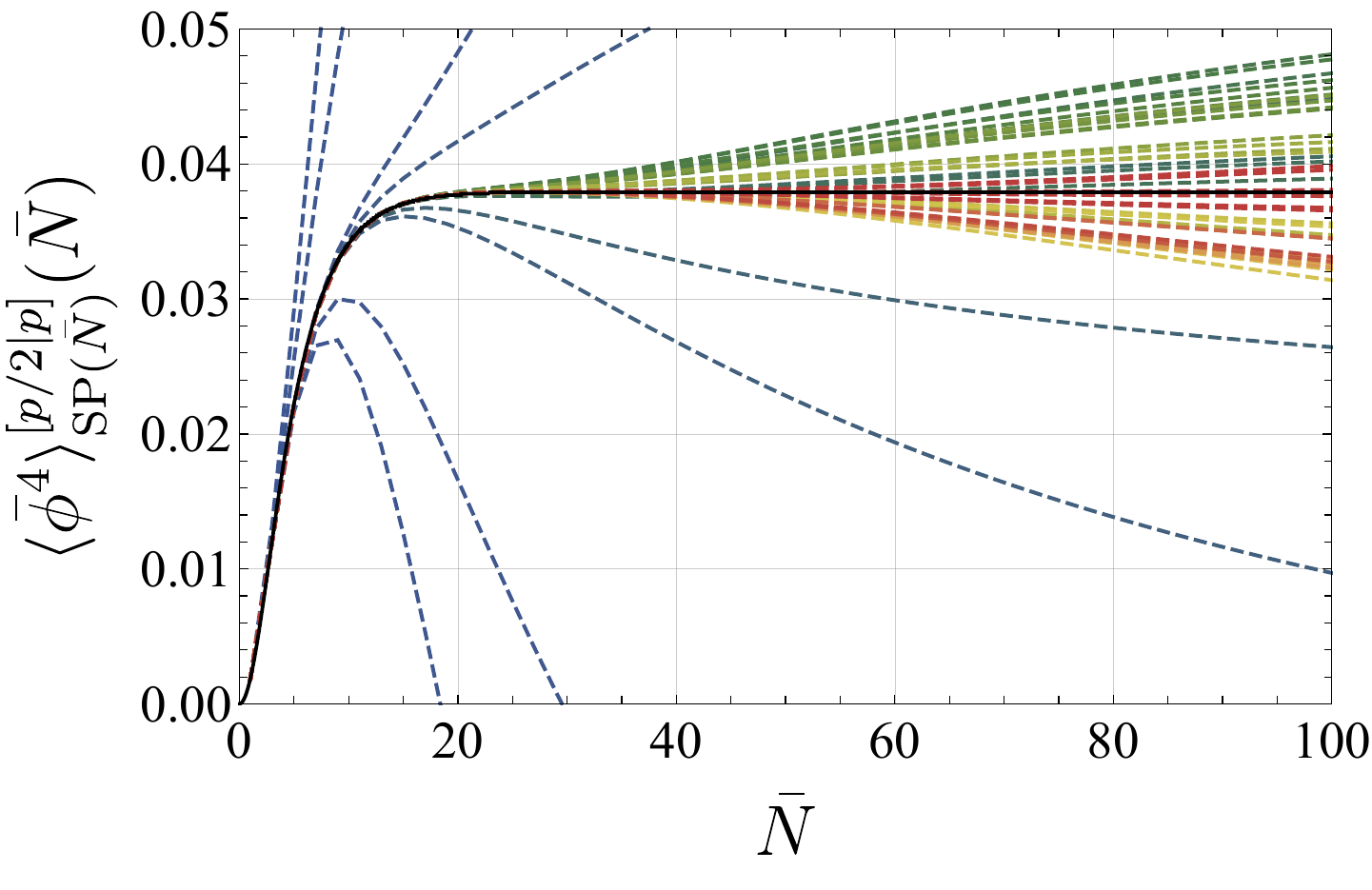}
  \end{minipage}
  \caption{
    Borel--Pad\'{e} summation $\expval{\bar{\phi}^2}_{\mathrm{SP} (\bar{N})}^{[p/2 | p]} (\bar{N})$ (\textit{left}) and $\expval{\bar{\phi}^4}_{\mathrm{SP} (\bar{N})}^{[p/2 | p]} (\bar{N})$ (\textit{right}) for $V = \lambda \phi^4 / 4$ with $\bar{N}$ being the expansion parameter.
    The orders are $p = 2, \, 4, \, \dots, \, 100$ and $p$ increases from blue to red.
  }
  \label{fig:Quartic_BorelPade_N}
\end{figure}

To avoid this issue, one may consider increasing the hierarchy between the orders of the numerator and denominator in the Pad\'{e} approximant, as the high peaks arise from insufficient suppression of the Borel--Pad\'{e} transformation for large $t$.
In Fig.~\ref{fig:Quartic_BorelPadeTransf_N} we show the Borel--Pad\'{e} transformation using $[m | n] = [p/2 | p]$,
\begin{align}
  \expval{ \bar{\phi}^2 }_{\mathrm{BP} (\bar{N})}^{[p/2 | p]} (t) 
  &= \qty[ \sum_{k = 0}^{\infty} \frac{\bar{a}_{2, k + 1}}{k!} t^k ]_{\mathrm{P}}^{[p/2 | p]} \,\, ,
  \\
  \expval{ \bar{\phi}^4 }_{\mathrm{BP} (\bar{N})}^{[p/2 | p]} (t) 
  &= \qty[ 
  \sum_{k = 0}^{\infty} \frac{\bar{a}_{4, k + 1}}{k!} t^k 
  ]_{\mathrm{P}}^{[p/2 | p]} \,\, .
\end{align}
The high peaks now disappear.
The corresponding Borel--Pad\'{e} resummation,
\begin{align}
  \expval{ \bar{ \phi }^2 }_{\mathrm{SP} (\bar{N})}^{[p/2 | p]} ( \bar{N})
  = \int_{0}^{\infty} \dd t \, e^{-t / \bar{N}} \expval{ \bar{ \phi }^2 }_{\mathrm{BP} (\bar{N})}^{[p/2 | p]} (t) \,\, ,
  \\
  \expval{ \bar{ \phi }^4 }_{\mathrm{SP} (\bar{N})}^{[p/2 | p]} ( \bar{N})
  = \int_{0}^{\infty} \dd t \, e^{-t / \bar{N}} \expval{ \bar{ \phi }^4 }_{\mathrm{BP} (\bar{N})}^{[p/2 | p]} (t) \,\, ,
\end{align}
is plotted in Fig.~\ref{fig:Quartic_BorelPade_N}.
We see that the lines nicely reproduce the exact result.

\section{Numerical precision in Borel--Pad\'{e} transformation}
\label{app:appendix2}

\begin{figure}
\centering
\includegraphics[clip, width = 0.95\linewidth]{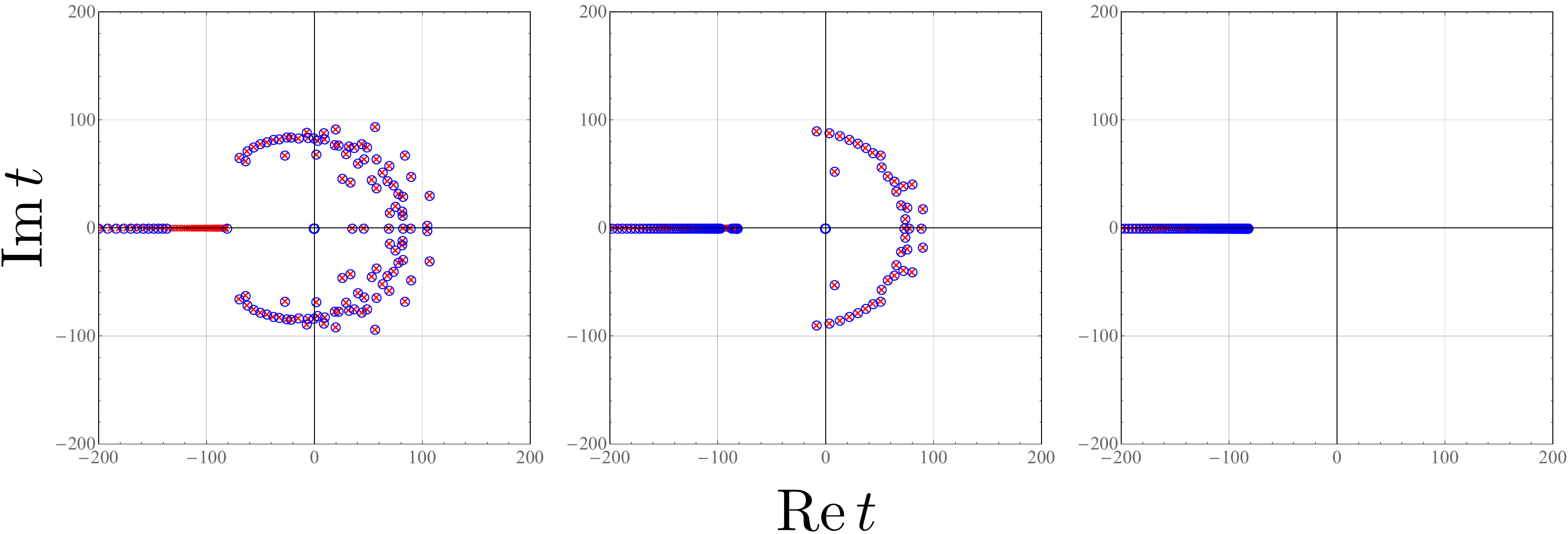}
\\[2.0ex]
\includegraphics[clip, width = 0.95\linewidth]{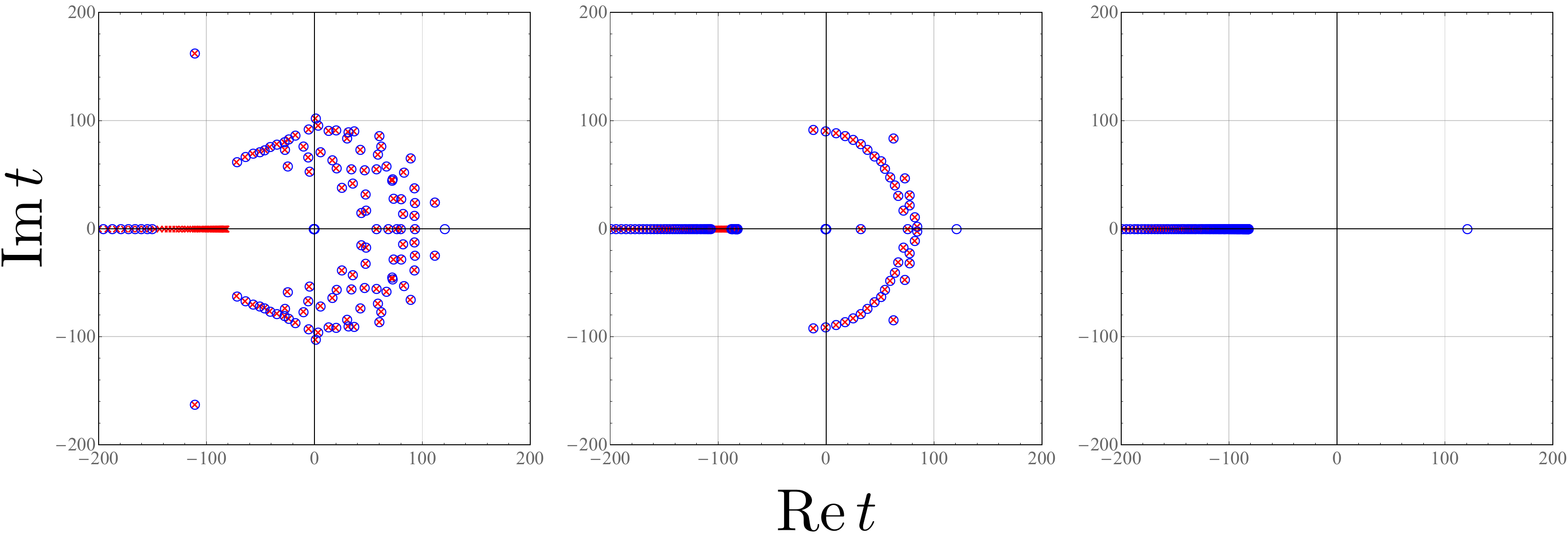}
\caption{
    Precision dependence of the distribution of the poles and zeros in the Borel--Pad\'{e} transformation $\expval{\bar{\phi}^2}_{\mathrm{BP} (\bar{N}^2)}^{[p - 1 | p + 1]} (t)$ (\textit{top}) and $\expval{\bar{\phi}^4}_{\mathrm{BP} (\bar{N}^2)}^{[p - 1 | p + 1]} (t)$ (\textit{bottom}) for $V = \lambda \phi^4 / 4$. 
    In all the panels, the order $p$ of the Borel--Pad\'{e} transformation is fixed to $p = 200$, and the numerical precision is changed as $100$ (\textit{left}), $200$ (\textit{middle}), and $300$ (\textit{right}). 
    The cluster along the horizontal axis implies a branch cut (with the poles and zeros appearing alternately), while the one along the circle for lower precisions indicates ghost pairs (with the poles and zeros appearing at the same locations). 
}
\label{fig:sing_singularity_precision}
\end{figure}

As mentioned in Sec.~\ref{subsec:sing_Borel_plane}, special care is needed when identifying the location of the poles and zeros in the Borel plane.
Figure~\ref{fig:sing_singularity_precision} shows how the location of the poles and zeros change depending on the numerical precision.
Note that the plot range is totally different from the main text: Figure~\ref{fig:sing_singularity_precision} corresponds to a zoom-in of Fig.~\ref{fig:sing_singularity} around the origin, calculated with different numerical precisions. 
In this figure, the order of the Borel--Pad\'{e} transformation is fixed to $p = 200$, and the precision is changed as $100$, $200$, and $300$ from left to right.
For precision below some threshold, poles and zeros start to appear along the circle of convergence $\abs{t} = \abs{t_0} \simeq 80$.
These poles and zeros appear in pairs at the same locations, and they are called \textit{zero-pole ghost pairs}~\cite{yamada2014numerical} (see also a recent progress \cite{Costin:2022hgc}).
This property helps to identify them as numerical artifacts, and indeed they disappear as the precision increases.

\clearpage

\bibliographystyle{JHEP}
\bibliography{reference}

\end{document}